# Cleaner air strengthens the Pacific trade winds without a detectable change in El Niño

Ioana Colfescu[1,2,3]*, Alice Pigelet[3], Chen Lu[1,2,3], Massimo A. Bollasina[4], Silvio Gualdi[5]

[1]National Centre for Atmospheric Science, University of St Andrews, St Andrews, UK

[2]School of Earth and Environmental Sciences, University of St Andrews, St Andrews, UK

[3]School of Mathematics and Statistics, University of St Andrews, St Andrews, UK

[4]School of GeoSciences, University of Edinburgh, Edinburgh, UK

[5]Fondazione Centro Euro-Mediterraneo sui Cambiamenti Climatici (CMCC), Bologna, Italy

*e-mail: ioana.colfescu@ncas.ac.uk

## Abstract

Air-quality policy will cut aerosol and ozone-precursor emissions, and studies disagree on how the tropical Pacific circulation and the El Niño–Southern Oscillation (ENSO) will respond. Here we show, in 12 Earth system models with and without air-quality controls, that the cuts strengthen the Pacific trade winds and offset half to two-thirds of the weakening projected without them. The response is absent with prescribed sea-surface temperatures and detectable only after 2040. It is consistent with the Indian Ocean warming relative to the Pacific, which also ranks the responses to cuts made region by region. ENSO amplitude changes by less than a tenth of its value, the smallest change this ensemble could detect. The change implied by the mean-state shift is a sixth of that, and an ENSO model trained without the reduced-emission runs predicts none. El Niño's regional impacts still change, because the background climate does.

## Main

The tropical Pacific sets much of the year-to-year variability of the global climate. Its trade winds and the Walker circulation, the overturning cell that rises over the warm pool and sinks over the eastern Pacific, control where tropical rain falls. The El Niño–Southern Oscillation (ENSO) that grows out of this coupled system shifts rainfall and temperature on every inhabited continent. How ENSO will respond to greenhouse warming is still debated[1]. Any forcing that changes the Pacific mean state therefore changes regional climate risk, and the coming air-quality cuts are one.

Emissions of near-term climate forcers (NTCFs), that is aerosols and the precursors of tropospheric ozone, are expected to fall steeply as countries act on air quality, with the largest cuts in Asia[2]. Aerosols weakened the South Asian summer monsoon over the twentieth century[3], and the recent East Asian clean-up has probably added to global and North Pacific warming[4]. Future cuts are projected to increase Asian monsoon rainfall[5] and to weaken the Northern Hemisphere Hadley cell further[6].

What the cuts do to the Pacific circulation is contested, and observed and simulated Walker trends already disagree[7,8]. Aerosols have been credited with part of the 1991–2010 strengthening of the trades[9], with delaying the warming of the eastern equatorial Pacific[10] and with steepening the zonal temperature gradient[11]. The shift of emissions from North America and Europe to Asia has been credited with strengthening the Walker circulation[12]. These results imply that removing aerosols would weaken the circulation, as regional emission-perturbation experiments found[13], although idealised Asian aerosol forcing did not reach the equatorial trades[14]. Hwang et al.[15] removed Northern Hemisphere aerosols in one model and found a fast weakening, then a slower, ocean-mediated strengthening[16]. Jeon et al.[17] found that a global aerosol reduction in the Regional Aerosol Model Intercomparison Project (RAMIP)[18] leaves the circulation unchanged on average. None examined ENSO or used the multi-model experiments designed for the question.

We test the question in those experiments, the Aerosol Chemistry Model Intercomparison Project (AerChemMIP) scenarios of the sixth Coupled Model Intercomparison Project (CMIP6)[19]. The ssp370 simulations follow the Shared Socioeconomic Pathway SSP3-7.0 of rising greenhouse gases and high aerosol emissions. The ssp370-lowNTCF simulations follow the same greenhouse-gas pathway with strong air-quality controls, so the difference between them isolates the climate effect of the NTCF cuts. We analyse 12 models with 2 to 9 members each (Extended Data Table 1), and each model counts once in every statistical test. The single-forcing experiments of the Detection and Attribution Model Intercomparison Project (DAMIP)[20] and the regional cuts of RAMIP test the mechanism we propose independently.

## Results

### The trade winds strengthen as the forcing grows

By 2036–2054 the NTCF cuts strengthen the equatorial Pacific trade winds and steepen the surface pressure gradient of the Walker circulation (Fig. 1). The zonal wind stress over the western Pacific becomes more easterly by 1.3 mPa (millipascal), about 4% of its mean, in 11 of 12 models ($P = 0.001$ in an exact sign-flip test on the 12 model-level differences; Methods). In 11 of 12 models our Walker index, the sea-level-pressure difference between the eastern Pacific and the Indo-western Pacific, rises by 0.09 hPa ($P = 0.001$). Both tilts steepen as well, the sea surface by 9 mm (8 of 8 models with ocean output, $P = 0.008$) and the 20 °C isotherm, which marks the thermocline, by 1.6 m (9 of 9, $P = 0.004$).

These changes appear only once the forcing difference between the scenarios has grown. Over 2015–2035 neither the wind stress ($P = 0.77$), the pressure gradient ($P = 0.88$) nor the tilts ($P =$ 0.82 and 0.74) differs between the scenarios (Supplementary Fig. 1). The late-minus-early difference is itself significant for the wind stress and the pressure gradient (9 of 12 models, $P =$ 0.04 and 0.03). The late-window results survive the removal of any one model ($P \leq 0.002$ for winds and pressure, $P \leq 0.016$ for the tilts). They also survive merging the two GISS models into one (10 of 11, $P = 0.002$) and restriction to the seven models with a realistic ENSO seasonal cycle (Extended Data Table 2).

The response is seasonal, and it takes until the 2040s to clear the noise (Fig. 2). The trades and the pressure gradient strengthen from May to December and change little from February to April (Supplementary Fig. 2). The global-mean temperature difference between the scenarios is significant from 2025–2029 (0.05 K, $P = 0.02$), the pressure gradient only from the 2040s (0.19 hPa in 2045–2049, $P = 0.001$). We cannot read a physical lag from this, because a response with no lag would reach significance as late in half of our simulations (Extended Data Fig. 1). The stronger winds raise evaporation over the western Pacific by 0.8 W $m^{-2}$ (11 of 11 models, $P =$ 0.001), so the zonal SST gradient changes little (0.02 K, 8 of 12, $P = 0.09$). The response shows in the thermocline tilt instead.

The cuts oppose the change projected without them (Fig. 2d). Relative to 1974–2014, ssp370 weakens the western-Pacific trades by 1.9 mPa in 2036–2055 (2.5 mPa in the eight models with ocean output). The NTCF cuts offset 55% of this weakening in those eight models (95% interval 25–108%) and 70% in all 12 (39–179%), so the central estimate is half to two-thirds and the intervals do not exclude a full offset.

**An ocean-mediated response consistent with Indian Ocean warming**

The circulation change does not appear when the ocean cannot respond (Fig. 3a). In atmosphere-only experiments with identical sea-surface temperatures (SSTs) prescribed in both scenarios, the same cuts raise the surface shortwave radiation over South and East Asia by 2.5–8.0 W $m^{-2}$, as in the coupled runs of the same models. Yet the Walker index changes by +0.02, 0.00 and −0.03 hPa in BCC-ESM1, CNRM-ESM2-1 and UKESM1-0-LL. Their coupled versions give +0.01, +0.15 and +0.20 hPa (Extended Data Table 3). We can show this in only two models, because BCC-ESM1's coupled response is itself small.

The Indian Ocean warms 0.045 K more than the tropical Pacific under the cuts (11 of 12 models, P = 0.001), most strongly from August to December (Fig. 3b), the months in which the trades strengthen most. Luo et al.[21] showed that an Indian Ocean warmer relative to the Pacific drives easterly anomalies over the western Pacific, a link later named the inter-basin thermostat[22,23].

The historical aerosol-only experiments of DAMIP test the link across models, because there the aerosol pattern differs from model to model (Fig. 3c and Extended Data Table 4). Across 13 models, the change in western-Pacific wind stress between 1850–1899 and 1995–2014 follows the change in Indian Ocean minus Pacific temperature. The correlation is $r = -0.69$ ($P = 0.013$) once NorESM2-LM, which has the largest change in both, is set aside, and $r = -0.89$ with it (Spearman $\rho = -0.70$, $P = 0.008$, all 13).

Each model's own variability gives a second test, and it does not separate the route from ENSO. We regressed the trades on the Indian Ocean minus Pacific difference in each model's unforced variability, on 5-year means of the historical and ssp370 members, and applied the slope to the forced change in the difference. It predicts a strengthening of 1.01 mPa against the 1.34 mPa simulated (95% interval 0.43 to 1.67 mPa), with the right sign in 10 of 12 models (Fig. 3d). Its correlation with the trades is −0.53 in the raw series and no more than 0.07 at any lag from −6 to +6 months once Niño3.4, the central Pacific temperature index (Methods), is removed. With Niño3.4 regressed out, the same calculation predicts 0.14 mPa, a tenth of the simulated change (Extended Data Fig. 2). The central Pacific temperature measured relative to the tropical mean predicts 0.69 mPa on its own, about half of the simulated change (strengthening in 11 of 12 models). Adding the Indian Ocean minus Pacific difference to it adds 0.03 mPa.

The regional experiments of RAMIP[18] test the same idea with the aerosols removed from one region at a time (9 models with 10 members each, 2031–2050; Extended Data Fig. 2d). A global cut strengthens the trades in 7 of 9 models ($P = 0.03$) while leaving the Walker index unchanged ($P = 0.52$), as Jeon et al.[17] reported. The cut over Africa and the Middle East, which warms the Indian Ocean relative to the Pacific in 8 of 9 models ($P = 0.05$), strengthens the trades in 8 of 9 ($P = 0.03$). The East Asian cut, which cools it slightly, does not (6 of 9 weaken, $P = 0.23$). Across the 39 region–model pairs the two changes go together ($r = -0.76$, and $r = -0.74$ within models), although the pairs share models.

Historical aerosol forcing cools the globe and steepens the zonal temperature gradient[10,11], but shifts neither the Indian Ocean relative to the Pacific on average ($P = 0.80$) nor the Walker circulation consistently ($P = 0.20$; Extended Data Table 4).

The rising branch of the circulation also shifts west. From May to December the Indian Ocean and Maritime Continent become wetter (11 of 12 models, $P = 0.001$) and the central Pacific drier (11 of 12, $P = 0.015$). The centre of the equatorial rain belt moves west (8 of 12, $P = 0.02$), and the ascent does the same in the three models with that field (Extended Data Fig. 3).

**No detectable change in ENSO, as the mean state implies**

The change in the mean state does not carry over to ENSO in any way we can detect (Fig. 4). We fitted a linear inverse model[24], a linear stochastic model of the coupled system, and a recharge oscillator[25] in temperature and thermocline depth to the SST, thermocline depth and wind stress of each member. We validated the fits by simulation (Extended Data Fig. 4). None of the growth rate, feedback coefficients or noise amplitudes differs between the scenarios (all $P > 0.1$; Fig. 4a). ENSO amplitude, the standard deviation of Niño3.4 SST, changes by −0.014 K in the eight models with ocean output (95% interval −0.090 to +0.073 K). Any change is therefore smaller than about 0.09 K, 8% of the amplitude, and the smallest change the ensemble could detect with 80% power is 0.116 K. The same test does detect the greenhouse-driven change. Between 1880–1899 and 2036–2055 the Niño3.4 standard deviation rises by 0.195 K in the 12 models with that index (11 of 12, $P = 0.006$), above the 0.17 K threshold for that set. The cuts change it by 0.011

K in the same metric and models (Fig. 4c). Two models change by more than the threshold on their own, BCC-ESM1 by −0.09 K and UKESM1-0-LL by +0.18 K.

Within each model, ENSO amplitude also covaries with the mean trades and thermocline tilt across members and periods. That sensitivity times the forced mean-state change predicts −0.018 K, 15% of the detection threshold and close to the −0.014 K simulated (Fig. 4d). Niño3.4 skewness is 0.19 higher under the cuts (7 of 8, P = 0.03), but as high in 2015–2035 as in 2036–2055, so unlike the mean-state signals it does not grow with the forcing. Ren and Liu[26] found that historical aerosols raised ENSO variability, and we see no sign of the reverse.

To separate changes in ENSO's dynamics from changes in its noise we also fitted a neural stochastic differential equation (SDE). It keeps the recharge-oscillator skeleton and adds a neural network for the nonlinear drift and for state-dependent noise. It is given a forcing index, which rises from zero in 2015 to one in 2054 under the cuts, without being told which variables respond. It learns an easterly forcing of the trade winds in all 8 models with ocean output (P = 0.008). Random relabelling of members between the scenarios does not reproduce this (P = 0.003 in a permutation test over the relabellings; Supplementary Table 1). Switching the learned forcing on in the drift alone changes ENSO amplitude by 0.01 K in the eight-model mean (P = 0.81 against the relabelled fits; Fig. 4b). Knowing the forcing does not help it predict an unseen member (Extended Data Fig. 5 and Supplementary Table 1). The drift and noise shares are the same under other forcing indices and network sizes (Methods).

Machine-learning checks agree[27–29] (Extended Data Fig. 6 and Methods). Classifiers detect the cuts from the forced warming pattern but not from ENSO variability alone, and frozen forecast networks forecast ENSO about as well in both scenarios.

**How large a change ENSO would need**

We then asked how far the background would have to move before ENSO responded, and whether its response could have been predicted without seeing it (Fig. 5). On the historical and ssp370 simulations only, we trained a hierarchy of models in which ENSO's parameters depend on the slowly varying background, from a linear inverse model through a physically constrained recharge oscillator to a neural SDE (Methods). The neural version raises the log-likelihood of the reduced-emission members it never saw by 0.16 nats (log-likelihood units) per month over a model with no background dependence (8 of 8 models, P = 0.008; Fig. 5a). It predicts them as well as an oracle, the same model trained also on them (P = 0.28). Within the range of backgrounds the models produce between 1850 and 2054, about 2.5 times the NTCF response, no level predicts a detectable change in ENSO amplitude (Fig. 5b). The crossings of the detection threshold are extrapolations beyond that envelope. They disagree, at 4, 16 and 32 times the response for the linear, neural and constrained levels (Fig. 5c). The constrained level, the only one that reproduces the greenhouse-driven rise when trained on 1850–2014 alone (0.27 K against 0.31 K simulated; Supplementary Fig. 3), crosses last.

**El Niño's impacts change through the climate it acts on**

Under the cuts El Niño's teleconnection patterns show no detectable change. One of 16 region-and-variable combinations differs in 2036–2054 and two of 16 in 2025–2054, about the rate expected by chance. The exception, the temperature response over north-eastern Brazil (P = 0.006), survives a false-discovery-rate adjustment over all 64 tests (q = 0.04, where q is the adjusted P) but not within the 16 teleconnection tests (Supplementary Table 2). The background does change, in several regions by as much as the El Niño signal (Fig. 6). That background change is the monsoon and warming response Wilcox et al.[5] reported. No change in the teleconnection accompanies it, so the background change adds to El Niño's impacts and does not reshape them, as Trascasa-Castro et al.[30] found under greenhouse warming.

The temperature changes are the secure part. The background warms by 0.12 to 0.29 K over India, the Maritime Continent, East Africa and the southern United States, and these four changes survive a false-discovery-rate adjustment over all 64 tests (q = 0.02 to 0.03; Supplementary Table 2). Over north-eastern Brazil the chance that an El Niño season falls in the hottest third of the ssp370 distribution rises from 0.75 to 0.96 (P = 0.008; Fig. 6c). For rainfall the evidence is suggestive only. The cuts raise the Indian summer background rainfall by 0.19 mm per day, a third of the El Niño deficit of 0.56 mm per day. Over southern Africa they offset about half of the El Niño drying (0.14 against 0.28 mm per day). Both have P = 0.02 but q = 0.09 and 0.08 after the adjustment, and both shrink in the longer 2025–2054 window. All the estimates rest on a few El Niño seasons per member.

## Discussion

In these models, air-quality policy strengthens the Pacific trade winds by mid-century and offsets half to two-thirds of the weakening projected without it. The three models with prescribed SSTs show none. It becomes detectable only after 2040 and is consistent with the warming of the Indian Ocean relative to the Pacific. Within each model that relation is largely ENSO's, however, and we have not shown the warming to cause the response. In observations since 1980, with ENSO and trends removed, the two covary in the HadISST and ESA-CCI sea-surface temperature datasets ($r = -0.41$, $P = 0.007$, and $r = -0.33$, $P = 0.07$; Supplementary Table 3).

If the Indian Ocean sets the sign, the earlier results do not conflict, because the sign depends on where the aerosols change. Ying et al.[12] found that the historical shift of emissions towards Asia strengthened the circulation, and a cut that removes the Asian aerosols is not the reverse of that shift, because it warms the Indian Ocean first. Jeon et al.[17] found no consistent Walker response to a global aerosol reduction in RAMIP, and nor do we in the same runs. The trades nevertheless strengthen there, and the regional cuts rank by what they do to the Indian Ocean. Their index measures ascent over 120–160°E, where the rising branch weakens as it shifts west (Extended Data Fig. 3).

Only 8 or 9 models published ocean output for the reduced-emission scenario, members are few, and the detection threshold for ENSO amplitude is about a tenth of the amplitude. The test finds the 0.2 K greenhouse-driven change but not the 0.08 K part of it since 1995–2014, so a change of that size would be missed. The prescribed-SST test and the separation of aerosols from ozone precursors rest on one to three models (Extended Data Table 3). The across-model relation in the historical experiments leans on one model with large changes, and the within-model test cannot be separated from ENSO.

ENSO shows no detectable response, and the size of the mean-state change appears to be the reason. In these models it is a damped oscillation kept going by weather noise[24,31], and a 4% change in the trades is too small to move either its damping or its noise appreciably. We therefore expect ENSO to change under these cuts by no more than about a tenth of its amplitude, whereas the climate on which El Niño acts does change. Forecast systems and impact studies that hold the background fixed would misstate the heat that comes with El Niño over India, the Maritime

Continent, East Africa, the southern United States and north-eastern Brazil. They may also misstate its rainfall deficits over India and southern Africa.

## Methods

### *Models, experiments and periods*

We use monthly output from the historical, ssp370 (SSP3-7.0) and ssp370-lowNTCF experiments of 12 AerChemMIP models (BCC-ESM1, CNRM-ESM2-1, EC-Earth3-AerChem, GFDL-ESM4, GISS-E2-1-G, GISS-E2-1-H, IPSL-CM5A2-INCA, MIROC6, MPI-ESM-1-2-HAM, MRI-ESM2-0, NorESM2-LM and UKESM1-0-LL; Extended Data Table 1). Atmospheric fields are available for all 12. Ocean fields are available for 8 or 9, because GFDL-ESM4, MIROC6 and UKESM1-0-LL published no ocean output for ssp370-lowNTCF. Members are matched across experiments by variant label, and each model contributes 2 to 9 members. Atmospheric fields are regridded to 5° by area-weighted averaging. The western and eastern equatorial boxes are 5°S–5°N, 120°E–180° and 5°S–5°N, 160°W–80°W. The Walker index is the sea-level-pressure difference between 5°S–5°N, 160°W–80°W and 5°S–5°N, 80°E–160°E. Its western box reaches into the eastern Indian Ocean. Every wind-stress quantity, including the western-Pacific wind stress of Figs 1–3, the within-model test and the DAMIP relation, is the zonal wind stress over 5°S–5°N, 160°E–150°W (the trade-wind index), which lies entirely within the Pacific. Negative wind-stress changes denote stronger easterlies and are quoted as magnitudes in the text. ENSO amplitude is the standard deviation of monthly Niño3.4 (5°S–5°N, 170°W–120°W) SST anomalies and persistence is their autocorrelation at lags of 6 and 12 months. GISS-E2-1-G and GISS-E2-1-H share an atmosphere and are treated as separate models; the leave-one-out tests bound the effect of either, and Extended Data Table 2 repeats the main tests with the two merged into one model ($n = 11$). The Indian Ocean and Pacific boxes for near-surface temperature are 20°S–20°N, 50–100°E and 20°S–20°N, 150°E–90°W. The early and late windows are 2015–2035 and 2036–2054 (2036–2055 for ocean fields).

### *Statistics and reproducibility*

Models are the unit of inference. For each quantity we average members within each model and experiment and form the paired difference between scenarios. We test the multi-model mean with an exact two-sided sign-flip permutation test over the $2^n$ sign assignments, where n is the number of models. The smallest attainable two-sided P is therefore 0.0005 for 12 models (all signs agreeing) and 0.008 for 8, and the smallest attainable with 11 of 12 agreeing is about 0.001. Confidence intervals come from a two-level bootstrap (4,000 resamples) that resamples models and then members within each model. We require a forced signal to be larger in 2036–2054 than in 2015–2035 and test the late-minus-early difference directly with the same sign-flip test. We repeat each test leaving one model out, and with the three physics versions of GISS-E2-1-G treated as one model or as separate models (Extended Data Table 2). That table also lists the early- and late-window results for six mean-state quantities. Member-level tests are not used for significance, because in this ensemble they gave confidence intervals 1.4 to 4.3 times too narrow. Cross-model relations are reported as Pearson r with Spearman ρ and a leave-one-out range where a single model carries weight. No data were excluded. Model sets are listed in Extended Data Table 1.

### *Fixed-SST and single-forcer experiments*

We use ssp370SST and its -lowNTCF, -lowAer, -lowBC, -lowO3 and -lowCH4 counterparts, atmosphere-only runs that prescribe identical SSTs and sea ice in each pair. Only BCC-ESM1, CNRM-ESM2-1 and UKESM1-0-LL published the circulation fields needed, with one member each, and only UKESM1-0-LL published the single-forcer runs. Annual paired differences over 2030–2049 are tested with a two-sided t-test whose sample size is reduced for lag-1 autocorrelation. The share of the NTCF surface shortwave change reproduced by each single-forcer cut is the slope of its regional pattern regressed on the lowNTCF pattern (Extended Data Table 3).

### *DAMIP experiments*

We use hist-aer and hist-GHG[20] from 13 and 14 models, up to five members each. Changes are formed between 1850–1899, 1965–1984 and 1995–2014 (Extended Data Table 4). Across-model relations use the member-mean change of each model.

### *RAMIP experiments*

We use the RAMIP[18] global (ssp370-126aer) and regional (Africa and the Middle East, East Asia, South Asia, North America and Europe) aerosol-reduction experiments. They come from the 9 models with output on CEDA, with 10 members each, and we form 2031–2050 differences from ssp370. RAMIP publishes no wind stress, so the trade-wind index there is the 10-m zonal wind over 5°S–5°N, 160°E–150°W; the Walker index and the Indian Ocean minus Pacific temperature are defined as above. Not every model ran every regional experiment, which leaves 39 region–model pairs. Tests use models as the unit, and the pairs that enter the across-pair correlation share models, so no P value is attached to it.

### *Vertical velocity*

BCC-ESM1, CNRM-ESM2-1 and UKESM1-0-LL archived 500 hPa vertical velocity for both scenarios. Equatorial (5°S–5°N) profiles are averaged over 2036–2054 and the change is evaluated over 80–130°E, 120–160°E and 160°W–100°W.

### *Dynamical hierarchy*

The state vector is $x = (T, h, \tau)$: Niño3.4 sea-surface temperature, sea-surface height averaged over the western box as a proxy for thermocline depth, and the trade-wind index. Anomalies are taken from each member's own monthly climatology and linear trend. The linear inverse model assumes $dx/dt = Lx + \xi$. We estimate $L = \log G(s)/s$ from the lag-covariance propagator $G(s)$ at $s = 2$ months and obtain the noise covariance from the fluctuation–dissipation relation. The recharge oscillator is the same estimator restricted to (T, h), with coefficients read in the form of Jin[25]. Fitted operators are validated by simulation against each model's ENSO amplitude, persistence and spectrum. An earlier one-step estimator with heat content averaged over the whole equatorial Pacific gave a negative Bjerknes feedback coefficient in 94% of members. The western box restores the recharge phase lead and a positive Bjerknes feedback, while the net growth rate stays negative, as it should for a damped oscillation sustained by noise (Extended Data Fig. 4).

### *Neural SDE*

The monthly change of the state is $x(t+1) - x(t) = f(x, s, F) + \sigma(x, s, F)\varepsilon$, where s is the calendar month, F the forcing index and ε standard Gaussian noise. The drift is $f = \mu_0 + \mu_F F + (A_0 + A_s(s) + A_F F)x - \delta x^3 + g(x, s, F)$, with cubic damping $\delta \geq 0$ and g a two-layer network of 16 units. The noise is $\log \sigma = c_0 + c_F F + c_s(s) + k(x, F)$. F is zero in ssp370. In ssp370-lowNTCF it rises linearly from zero in 2015 to one in 2054, which approximates the growth of the global-mean temperature difference between the scenarios. Anomalies of both experiments are taken from the ssp370 ensemble-mean climatology and trend of the same model, so that the reduced-emission members keep their forced departure. We fit one SDE per model by maximising the one-step Gaussian likelihood over 2015–2054, with weight decay on the network terms, and refit it 20 times on member-bootstrap samples. We fit a linear version without g and k in the same way. We attribute pathways by simulating each fit for 400 years in 40 chains with each forcing term switched on separately, using common random numbers. In the placebo test we reassign members between the experiments within each model, keeping the number in each experiment fixed. We fit every distinct assignment (20 for models with three members per experiment and 6 for those with two) and 200 random assignments for GISS-E2-1-G, which has nine, 313 fits in all. We compare the eight-model mean of the real fits with 200,000 eight-model means formed by drawing one placebo fit per model, and P = (1 + number of draws at least as extreme)/(1 + 200,000), two-sided (Supplementary Table 1). In the held-out test we remove one ssp370-lowNTCF member and compare SDEs that do and do not see F on it. We repeated the pathway split with four forcing indices (the linear ramp, the global-mean temperature difference, a step and a quadratic) and four architectures (8 and 32 units, no noise network, and the linear version). The drift pathway stayed between −0.02 and +0.01 K and the noise pathway near +0.03 K in all of them. A weakly regularised variant did not learn the forcing in 3 of 8 models and is not interpreted.

***Machine learning***

Classifiers take annual-mean maps of near-surface temperature, sea-level pressure and zonal wind stress over 60°S–60°N for 2025–2054. Each member is referenced to its own 2015–2024 mean. We train on 11 models and test on the twelfth, with equal weight per model, using L2-regularised logistic regression and a two-layer convolutional network (three seeds). The pattern task removes each map's area mean. The tropical-Pacific task uses 20°S–20°N, 120°E–80°W. The variability task uses 12-month equatorial Hovmöller diagrams over 2036–2054 with each member's climatology and trend removed. Relevance is gradient × input. The forecast networks follow Lin et al.[29]: trained on the historical runs of the other models over 1900–1980, frozen, and applied to the held-out model's scenario runs over 2036–2055 with 20 seeds. The pattern classifier reaches 60% held-out accuracy, above chance in 12 of 12 models (P = 0.0005 in the sign-flip test), whereas the tropical-Pacific and variability classifiers average 0.50–0.51 (3 to 6 of 12 models above chance, P ≥ 0.45). The change in the forecast networks' anomaly correlation at leads of 12–23 months under the cuts is +0.04 (two-level 95% interval −0.06 to +0.13, n = 11 models).

### *Predicted ENSO change and detection threshold*

Each ENSO metric is regressed on the mean trade-wind stress and thermocline tilt across members, experiments and windows after removing each model's mean. The product of these sensitivities and the forced mean-state change is the predicted change. The detection threshold is 2.8 times the two-level standard error of the simulated change (80% power at 5% significance).

### *Background-conditioned emulator*

The background state is the 21-year running mean of the trade-wind index, the thermocline tilt, the Niño4 minus Niño3 temperature difference and the Niño3.4 temperature. ENSO anomalies are departures from the running climatology. Four models are fitted on historical (1850–2014) and ssp370 (2015–2054) data. The first has no background dependence. In the second, every operator entry and noise amplitude is linear in the background. In the third, the background acts only on the Bjerknes and thermocline feedbacks and on the noise of temperature and wind stress, with cubic damping. The fourth adds neural networks for nonlinear drift and noise to the third. The oracle is the neural version trained also on the other ssp370-lowNTCF members. Its likelihood gain over the neural emulator on the held-out member is 0.012 nats per month ($P = 0.28$, $n = 7$ models with two or more reduced-emission members). ENSO statistics come from 300-year simulations at fixed background with common random numbers. The training envelope is the largest multiple of the NTCF background change found anywhere in a model's 1850–2054 record.

### *Positive control*

We applied the detection test and the recharge-oscillator fits unchanged to the greenhouse-driven change in the same 12 models, between 1880–1899 in the historical runs and 2036–2055 in ssp370, and to the shorter step from 1995–2014. The metric is the standard deviation of Niño3.4, so that all 12 models can be used; the aerosol change in the same metric is reported alongside. We trained the physically constrained emulator (the third level of the hierarchy) on 1850–2014 only, in the eight models with ocean output. We then used it to predict the change in ENSO amplitude between the 1880–1899 background of the historical runs and the 2036–2054 background of ssp370. We trained and tested the other levels in the same way, and only the constrained level reproduces the simulated change (Supplementary Fig. 3). The whiskers in Supplementary Fig. 3b span five refits with members resampled. Here amplitude is the standard deviation of Niño3.4 anomalies about the 21-year running climatology, as in the emulator, and is compared with the same quantity in the simulations.

***Within-model Indian Ocean test***

For each model we took unforced variability from all historical (1850–2014) and ssp370 (2015–2054) members after subtracting the ensemble mean of each experiment at each month (a 31-year running mean for experiments with a single member). We then averaged it into non-overlapping 5-year means. We regressed the western-Pacific wind stress on the Indian Ocean minus Pacific temperature difference across these means and multiplied the slope by the forced change in the difference under the cuts over 2036–2054. We repeated the calculation with Niño3.4 regressed out of both series, with Indian Ocean temperature minus the 20°S–20°N mean as the predictor, and with Niño3.4 or Niño4 temperature as the predictor. Niño3.4 measured relative to the 20°S–20°N mean predicts 0.69 mPa, about half of the simulated change (strengthening in 11 of 12 models), and the Indian Ocean minus Pacific difference adds 0.03 mPa when both are used. The absolute Niño3.4 index carries the tropical-mean warming and predicts the wrong sign. Splitting the unforced data by El Niño, La Niña and neutral months recovers 0.36 to 0.58 of the simulated change in each phase (11 of 12 models). The lead–lag correlation between the unforced Indian Ocean minus Pacific difference and the trades is −0.53 and falls to 0.07 or less at every lag from −6 to +6 months once Niño3.4 is removed. Intervals for the 12-model mean come from a two-level bootstrap over models and, within each model, over 5-year means and members. Lead–lag relations use monthly values of the 12-model mean forced difference with its calendar-month mean removed; the block version uses 5-year means of each scenario member with model fixed effects.

***Emergence test***

The forcing is represented by the 12-model mean difference in global-mean temperature, smoothed by a quadratic in time that is zero before 2015, when the scenarios are identical. We regressed the annual 12-model mean difference in the Walker index (and in the trade-wind index) on this forcing lagged by 0 to 20 years. The best-fit lag and its 95% range come from 2,000 bootstrap resamples of the models. In a parallel simulation we added a response proportional to the unlagged forcing to each model's own residual noise (circular resampling of 5-year segments) and tested it block by block with the sign-flip test, 1,000 times. We report the fraction of simulations whose first significant block falls at or after the block observed.

### *El Niño impacts*

For each member and season, the teleconnection β is the regression of the seasonal anomaly of precipitation or temperature on seasonal Niño3.4. The impact of an El Niño of 1.5 K relative to the ssp370 climatology of the same model is $I = (m - m_{ssp370}) + 1.5\beta$, where m is the seasonal mean. The change in impact is the sum of a background term and a teleconnection term. Exceedance probabilities use El Niño seasons, defined as Niño3.4 above one standard deviation, and the terciles of the ssp370 distribution. The eight regions and seasons are shown in Fig. 6 and listed in Supplementary Table 2.

### *Data availability*

CMIP6 AerChemMIP and DAMIP output is available from the Earth System Grid Federation (https://esgf-node.llnl.gov/search/cmip6/) and the CEDA archive (https://catalogue.ceda.ac.uk/), and RAMIP output from the CEDA archive. HadISST 1.1 is available from the Met Office Hadley Centre (https://www.metoffice.gov.uk/hadobs/hadisst/), ESA-CCI SST v2.1 from the CEDA archive and ERA5 from the Copernicus Climate Data Store (https://cds.climate.copernicus.eu/). The indices and tables that underlie every figure will be deposited at Zenodo before publication and are available from the corresponding author on request.

### *Code availability*

The analysis code (Python) will be deposited at Zenodo before publication and is available from the corresponding author on request.

## Acknowledgements

We thank the modelling groups for the AerChemMIP, DAMIP and RAMIP simulations and the Earth System Grid Federation and CEDA for distributing them. This work used JASMIN, the UK collaborative data analysis facility. I.C. and C.L. were supported by the National Centre for Atmospheric Science and the University of St Andrews.

## Author contributions

I.C. conceived the study, designed and performed the analysis and wrote the paper. A.P. and C.L. contributed to the machine-learning analysis and the figures. M.A.B. and S.G. contributed to the interpretation of the aerosol experiments. All authors discussed the results and commented on the manuscript.

## Competing interests

The authors declare no competing interests.

## Figure legends

**Fig. 1 | Cleaner air strengthens the Pacific trade winds and the Walker pressure gradient.** a–c, Difference between ssp370-lowNTCF and ssp370 over 2036–2054, mean of 12 models, in near-surface air temperature minus its 20°S–20°N mean (a), sea-level pressure (b) and zonal wind stress (c). Dots mark grid points where at least 10 of 12 models agree in sign. Boxes mark the western (120°E–180°) and eastern (160°W–80°W) equatorial Pacific in a and b, and the wind-stress region (160°E–150°W) in c. d–g, Per-model differences in western-Pacific wind stress (d), the Walker sea-level-pressure index (e), sea-level tilt (f) and 20 °C isotherm tilt (g) in the early (2015–2035) and late (2036–2054; 2036–2055 for f, g) windows; each dot is one model's member mean ($n = 12$ models in d, e; $n = 8$ and 9 in f, g). Bars show the multi-model mean. Numbers give the models agreeing with the sign of the mean; asterisks mark $P < 0.05$ in an exact two-sided model-level sign-flip test.

**Fig. 2 | The response is seasonal, emerges late and opposes the projected weakening.** a,b, Calendar-month difference in western-Pacific wind stress (a) and the Walker index (b) over 2036–2054, mean of 12 models; orange bars are significant at $P < 0.05$ (sign-flip test, $n = 12$) and grey bars are not. c, Five-year means of the 12-model difference in the Walker index (top) and global-mean temperature (bottom), with two-level 95% intervals; filled symbols mark $P < 0.05$ and the first significant block is labelled. d, Western-Pacific wind-stress change under ssp370 (2036–2055 minus 1974–2014) and under the NTCF cuts (ssp370-lowNTCF minus ssp370, 2036–2055); dots are the 12 models and bars their mean.

**Fig. 3 | The response requires a coupled ocean and covaries with the Indian Ocean.** a, Walker index response to the NTCF cuts with prescribed sea-surface temperature (one member each, 2030–2049) and in the coupled runs of the same three models (2036–2054). b, Calendar-month difference under the cuts over 2036–2054, mean of 12 models, in surface shortwave radiation over South Asia, Indian Ocean minus Pacific near-surface temperature (20°S–20°N), the Walker index and the western-Pacific wind stress, each normalised by its largest monthly value. c, Western-Pacific wind-stress change against the change in Indian Ocean minus Pacific temperature, for the historical aerosol-only runs (1995–2014 minus 1850–1899, grey, $n = 13$ models) and the NTCF cuts (2036–2054, orange, $n = 12$ models, with CNRM-ESM2-1 and MRI-ESM2-0 overlapping). The grey line is the least-squares fit to the historical aerosol runs (r =

−0.89, P < 0.001) and the dashed line the fit without NorESM2-LM, the point at lower right (r = −0.69, P = 0.013). The NTCF responses have too little spread across models to show the relation on their own (r = −0.22). d, Forced change in western-Pacific wind stress under the cuts (2036–2054) predicted from each model's own unforced relation between the trades and the Indian Ocean minus Pacific temperature (orange) and the central Pacific (Niño3.4) temperature measured relative to the 20°S–20°N mean (grey), against the simulated change. The relative index predicts 0.69 mPa (right sign in 10 of 12 models) and the Indian Ocean minus Pacific difference 1.01 mPa (10 of 12), against 1.34 mPa simulated. Unforced relations use 5-year means of the historical and ssp370 members (n = 12 models); the line is 1:1.

**Fig. 4 | No detectable change in ENSO.** a, Change in fitted recharge-oscillator and linear-inverse-model quantities under the cuts over 2036–2055, as a percentage of the ssp370 value, with two-level bootstrap 95% intervals (n = 8 models). b, Change in ENSO amplitude attributed by the neural SDE (blue) and its linear version (grey) to the total, drift and noise pathways. Dots are the 8-model mean and bars 95% intervals, and shading marks the detection threshold (80% power). c, Positive control: change in Niño3.4 standard deviation in the same 12 models for the greenhouse-driven step (1880–1899 to 2036–2055, ssp370), the step since 1995–2014 (0.08 K, 8 of 12, P = 0.19), and the NTCF cuts (2036–2055), with two-level 95% intervals. The shaded band is the detection threshold, and numbers give models agreeing and P from the sign-flip test. d, Changes in four ENSO metrics (amplitude, growth rate, SST noise and Bjerknes feedback) predicted from the mean-state change (blue) and simulated (orange), in units of the detection threshold, with two-level 95% intervals (n = 8 models).

**Fig. 5 | ENSO under the cuts is predictable from the other forcings.** a, Likelihood gain on ssp370-lowNTCF members, in nats per month, for background-dependent models trained without them (linear, recharge oscillator, neural) relative to a model with no background dependence, and the gain of an oracle trained on the other ssp370-lowNTCF members over the neural emulator, for which positive values mean that the oracle does better. Dots are models (n = 8, and n = 7 for the oracle, which needs two or more reduced-emission members) and bars the mean. b, Emulated against simulated change in ENSO amplitude per model for the recharge oscillator and the neural emulator (n = 8 models). Shading marks the detection threshold, the line is 1:1 and r is the across-model correlation. c, Emulated change in ENSO amplitude as the NTCF background change is

scaled by 1 to 32 (n = 8 models). The dashed line is the detection threshold, and the dotted line and shading mark backgrounds beyond the 1850–2054 range of the median model (2.5 times).

**Fig. 6 | El Niño's impacts change through the background climate.** a,b, Precipitation (a) and temperature (b) impact of an El Niño of 1.5 K under ssp370 (grey) and its change under the NTCF cuts, split into a background term (orange) and a teleconnection term (blue), 2036–2054, mean of 12 models; asterisks mark $P < 0.05$ in the sign-flip test before adjustment for multiple tests (Supplementary Table 2). c, Probability that an El Niño season falls in the extreme tercile of the ssp370 distribution, under ssp370 (blue) and the NTCF cuts (orange), for 2036–2054 (circles) and 2025–2054 (squares).

## Extended Data

**Extended Data Fig. 1 | No detectable lag behind the forcing.** a, Five-year-block difference in the Walker index and in global-mean temperature between the scenarios, mean of 12 models, with two-level 95% intervals and the first significant block marked. b, Variance of the annual 12-model mean Walker-index and trade-wind differences explained by the lagged forcing as a function of lag, with the best fit for the Walker index (13 years), which explains less than 1% more variance than no lag, and its 95% range (0–18 years, shaded). c, Distribution of the first significant block for a simulated response with no lag, for the Walker index and the trades; triangles mark the observed first significant blocks, and 51% and 62% of simulations reach significance at or after them.

**Extended Data Fig. 2 | The Indian Ocean route within each model.** a, Regression slope of the western-Pacific wind stress on the Indian Ocean minus Pacific temperature difference (orange), on the same difference with Niño3.4 regressed out (purple), and on Indian Ocean temperature minus the tropical mean (green), in each model's unforced variability (5-year means of historical and ssp370 members), with 95% intervals (n = 12 models). b, Lead–lag correlation between the forced trade-wind and temperature-difference series (monthly 12-model mean difference, unsmoothed and as 12-month running means), with the peak lag and its bootstrap range. c, The same relation across members and 5-year blocks of the scenario runs with model fixed effects, with and without Niño3.4 removed; numbers give the models with a negative slope. The relation holds within the same block in 12 of 12 models and the Indian Ocean does not lead by a block. d, Change in the western-Pacific 10-m zonal wind in the RAMIP experiments (2031–2050 minus

ssp370; 9 models with 10 members each) for the global cut and the four regional cuts. Dots are models and bars their mean, and negative values are stronger easterlies. Numbers give the models in which the easterlies strengthen and P from the sign-flip test (n = 9); orange bars are significant at P < 0.05. Across the 39 region–model pairs the wind change follows the Indian Ocean minus Pacific change (r = −0.76).

**Extended Data Fig. 3 | The rising branch and the rain shift west.** a, Change in May–December precipitation under the cuts over 2036–2054, mean of 12 models; dots mark agreement of at least 10 of 12 models in sign. The Indian Ocean and Maritime Continent become wetter (11 of 12 models, P = 0.001) and the central Pacific drier (11 of 12, P = 0.015). b, Change in the longitude of the centre of the equatorial rain belt per model (n = 12; westward in 8 of 12, P = 0.02). c, Equatorial (5°S–5°N) 500 hPa vertical velocity under ssp370 over 2036–2054 for the three models with this field in both scenarios. d, Change under the NTCF cuts; positive values are more descent. Shading marks the boxes 80–130°E, 120–160°E and 160°W–100°W; two of the three models show weaker ascent over 120–160°E, the box of the Walker index of Jeon et al.[17], while all three show stronger ascent over 80–130°E.

**Extended Data Fig. 4 | Dynamical hierarchy.** a, Growth rate from the one-step estimator with full-basin and western heat-content boxes; the dashed line is the observed value. b, ENSO amplitude simulated by the fitted linear inverse models against the amplitude of the model that was fitted, for the historical, ssp370 and ssp370-lowNTCF windows. c, Change in fitted parameters and in Niño3.4 standard deviation and skewness for the full scenario window 2015–2055 (grey) and the late window 2036–2055 (orange), with two-level 95% intervals (n = 8 models).

**Extended Data Fig. 5 | Neural SDE.** a, Learned forcing coefficients (blue neural, grey linear) with 95% intervals, models agreeing and P. b, Learned against simulated mean-state response per model. c, Pathway attribution of the ENSO amplitude change; shading marks the detection threshold.

**Extended Data Fig. 6 | Machine-learning detection and forecast skill.** a, Held-out-model accuracy by year for each classification task and classifier. b, Per-model accuracy with models above chance and P (sign-flip test, n = 12). c,d, Relevance maps (gradient × input) for near-surface temperature and sea-level pressure. e,f, Anomaly correlation by lead for frozen forecast

networks in historical, ssp370 and ssp370-lowNTCF. g, Change in skill under the cuts at leads 12–23 months with two-level 95% intervals (n = 11 models).

**Extended Data Table 1 | Models and members.** Models, variant labels and member counts in each experiment (historical, ssp370, ssp370-lowNTCF, fixed-SST and single-forcer pairs, hist-aer, hist-GHG), the fields available for each, and which tests each model enters.

**Extended Data Table 2 | Sensitivity of the mean-state response.** Early- and late-window changes with models agreeing and P for six mean-state quantities (growth test); leave-one-model-out P values; GISS-E2-1-G physics-version variants and GISS-E2-1-G and GISS-E2-1-H merged into one model; the seven-model subset; seasonality by calendar month.

**Extended Data Table 3 | Fixed-SST and single-forcer responses.** Changes in surface shortwave radiation, Walker index and wind stress in the fixed-SST experiments of three models and the single-forcer experiments of UKESM1-0-LL, with the share of the lowNTCF shortwave pattern reproduced by each forcer (aerosols alone 90%; the black-carbon and ozone-precursor cuts weaken the Walker index by 0.08 and 0.06 hPa, $P = 0.002$ and 0.007) and P values from the autocorrelation-corrected t-test; also the coupled-run changes in the subtropical easterlies and in eastern-Pacific subsurface minus surface temperature (−0.03 K, 7 of 9 models, $P = 0.37$).

**Extended Data Table 4 | DAMIP epoch changes.** Walker index, wind stress, Indian Ocean minus Pacific temperature, zonal gradient and global-mean temperature changes between 1850–1899, 1965–1984 and 1995–2014 in hist-aer and hist-GHG, with models agreeing, P, and the leave-one-out range of the across-model correlation; hist-aer cools the globe by 0.63 K (13 of 13 models), steepens the zonal temperature gradient by 0.12 K (9 of 13, $P = 0.009$) and changes the Walker index by 0.13 hPa (7 of 13, $P = 0.20$), and hist-GHG weakens the Walker circulation in 14 of 14 models ($P < 0.001$).

## Supplementary Information

Supplementary Information (to accompany the submission). Supplementary Table 1 gives the neural SDE placebo (permutation over member relabellings) and held-out tests by model. Supplementary Table 2 gives the El Niño impact terms for both windows and all 16 region–variable combinations, with a false-discovery-rate adjustment. Supplementary Table 3 gives the observed relation between the Indian Ocean minus Pacific temperature and the western-Pacific zonal wind (HadISST and ESA-CCI SST with ERA5 1000-hPa wind, 1980–2017, annual and 5-year means, with and without Niño3.4 and trends removed). Supplementary Fig. 1 shows the growth test for six mean-state quantities, Supplementary Fig. 2 the seasonality of the mean-state response, and Supplementary Fig. 3 the emulator positive control per model.

# Figures

## Fig. 1

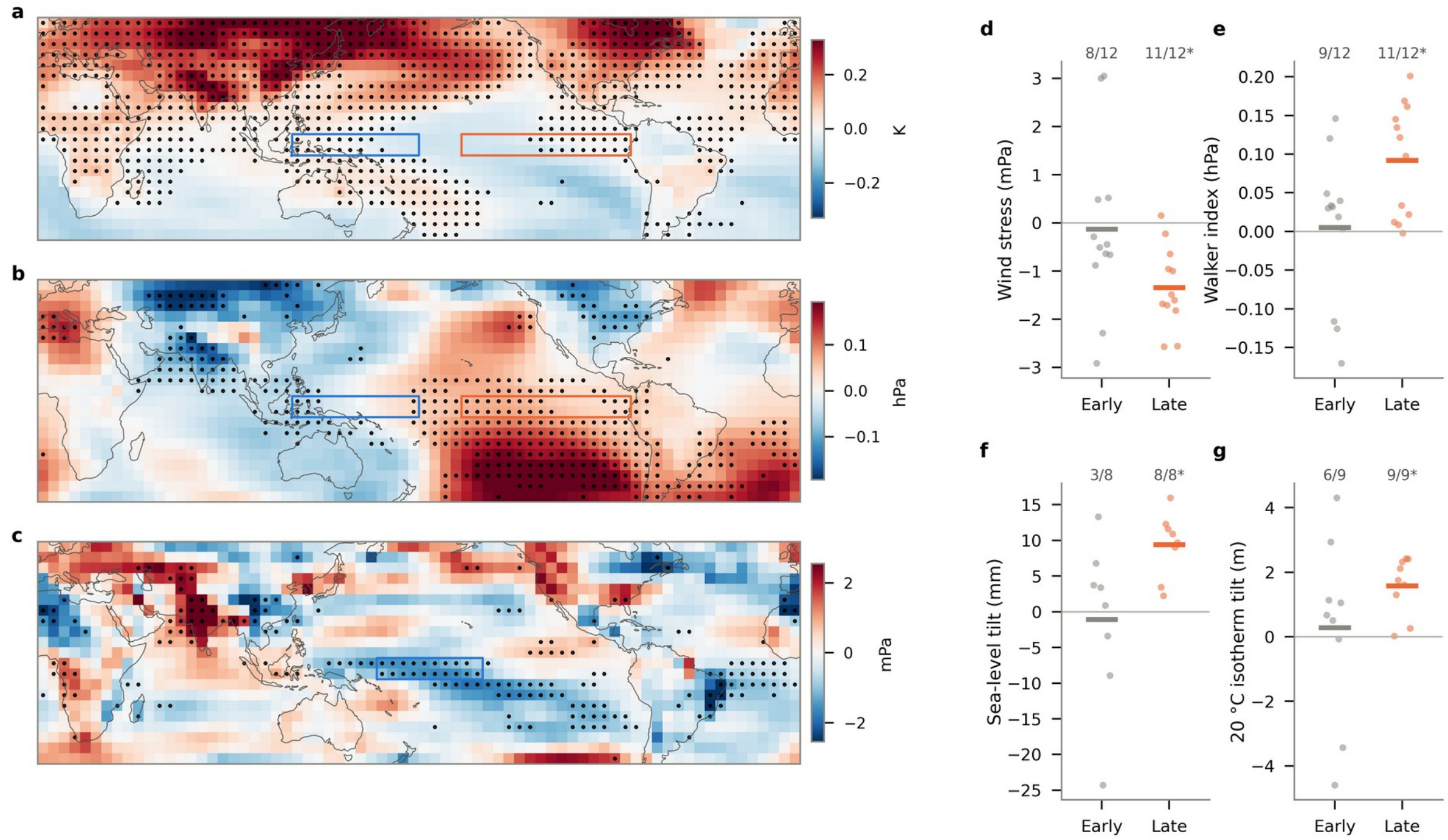

## Fig. 2

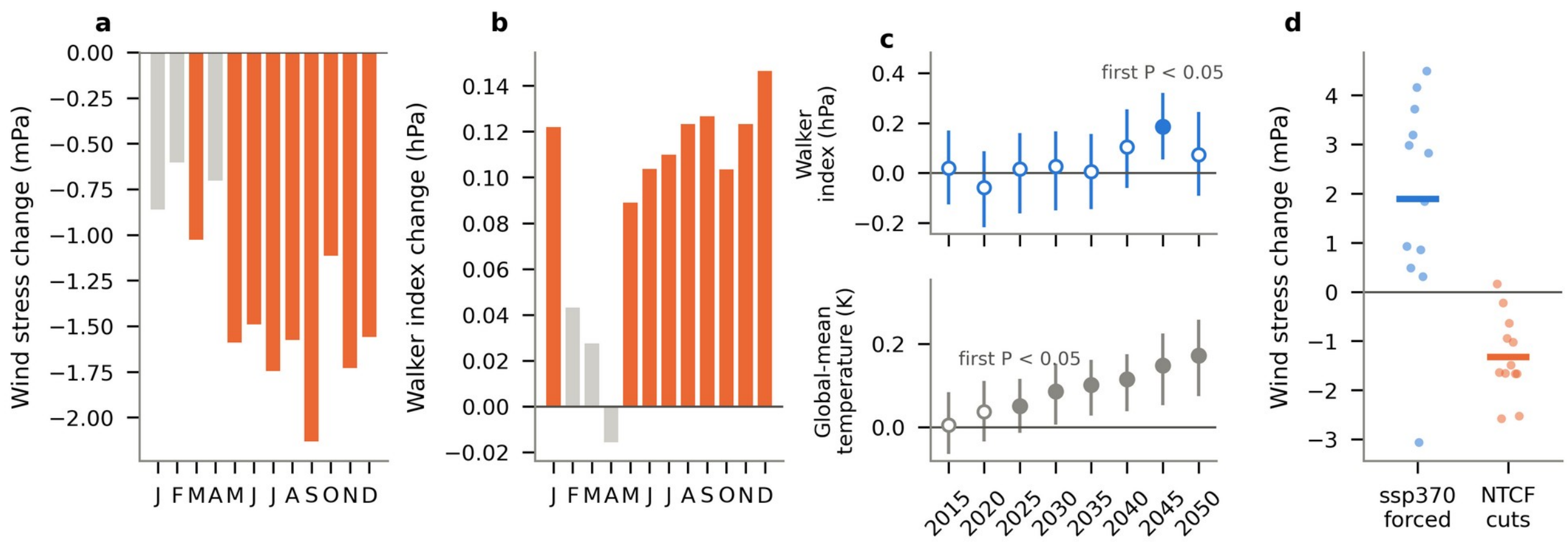

## Fig. 3

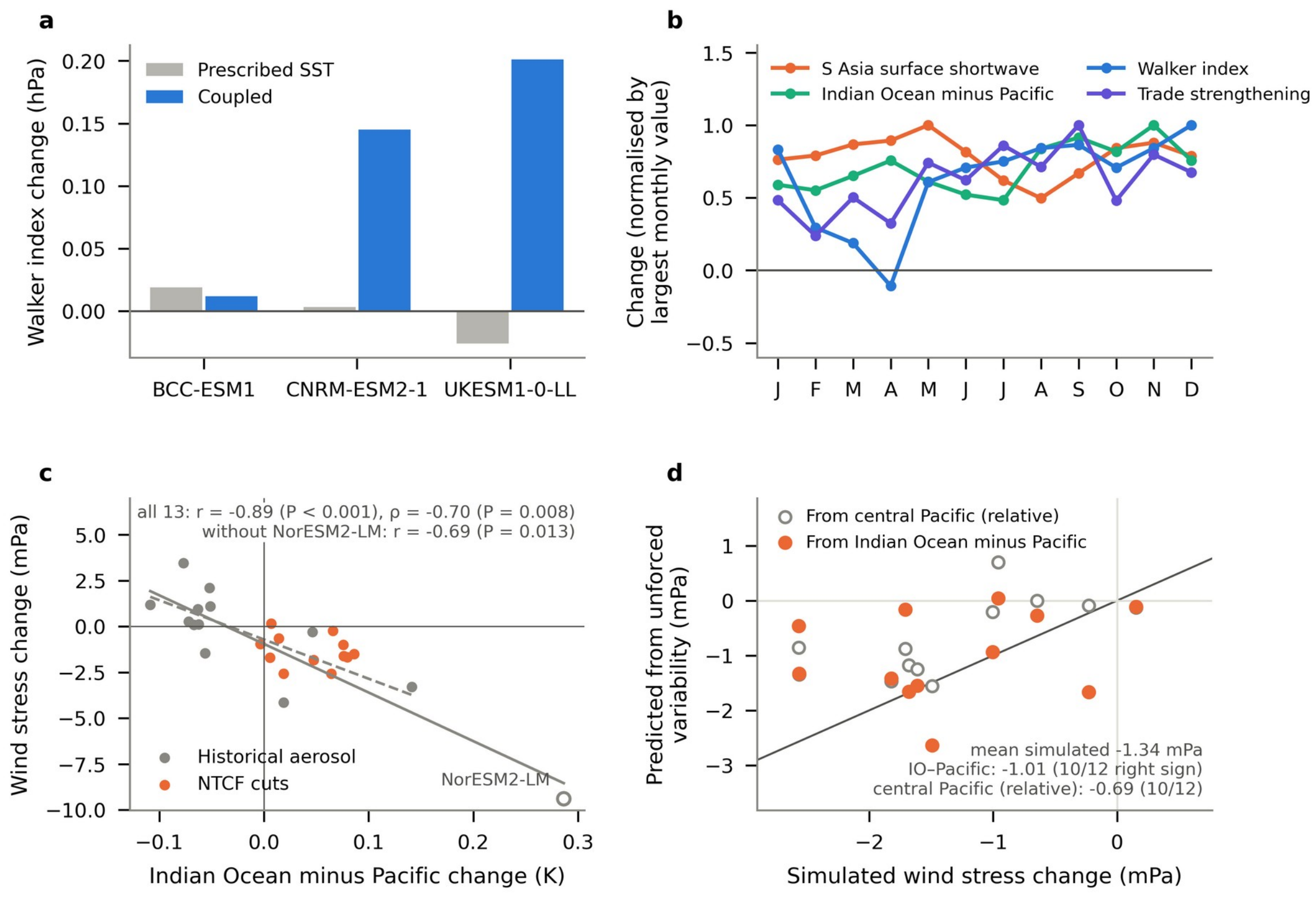
a
Walker index change (hPa)
Prescribed SST
Coupled
BCC-ESM1
CNRM-ESM2-1
UKESM1-0-LL
b
Change (normalised by largest monthly value)
S Asia surface shortwave
Indian Ocean minus Pacific
Walker index
Trade strengthening
J F M A M J J A S O N D
c
all 13: r = -0.89 (P < 0.001), ρ = -0.70 (P = 0.008)
without NorESM2-LM: r = -0.69 (P = 0.013)
Wind stress change (mPa)
Historical aerosol
NTCF cuts
NorESM2-LM
Indian Ocean minus Pacific change (K)
d
From central Pacific (relative)
From Indian Ocean minus Pacific
Predicted from unforced variability (mPa)
mean simulated -1.34 mPa
IO–Pacific: -1.01 (10/12 right sign)
central Pacific (relative): -0.69 (10/12)
Simulated wind stress change (mPa)

**Fig. 4**

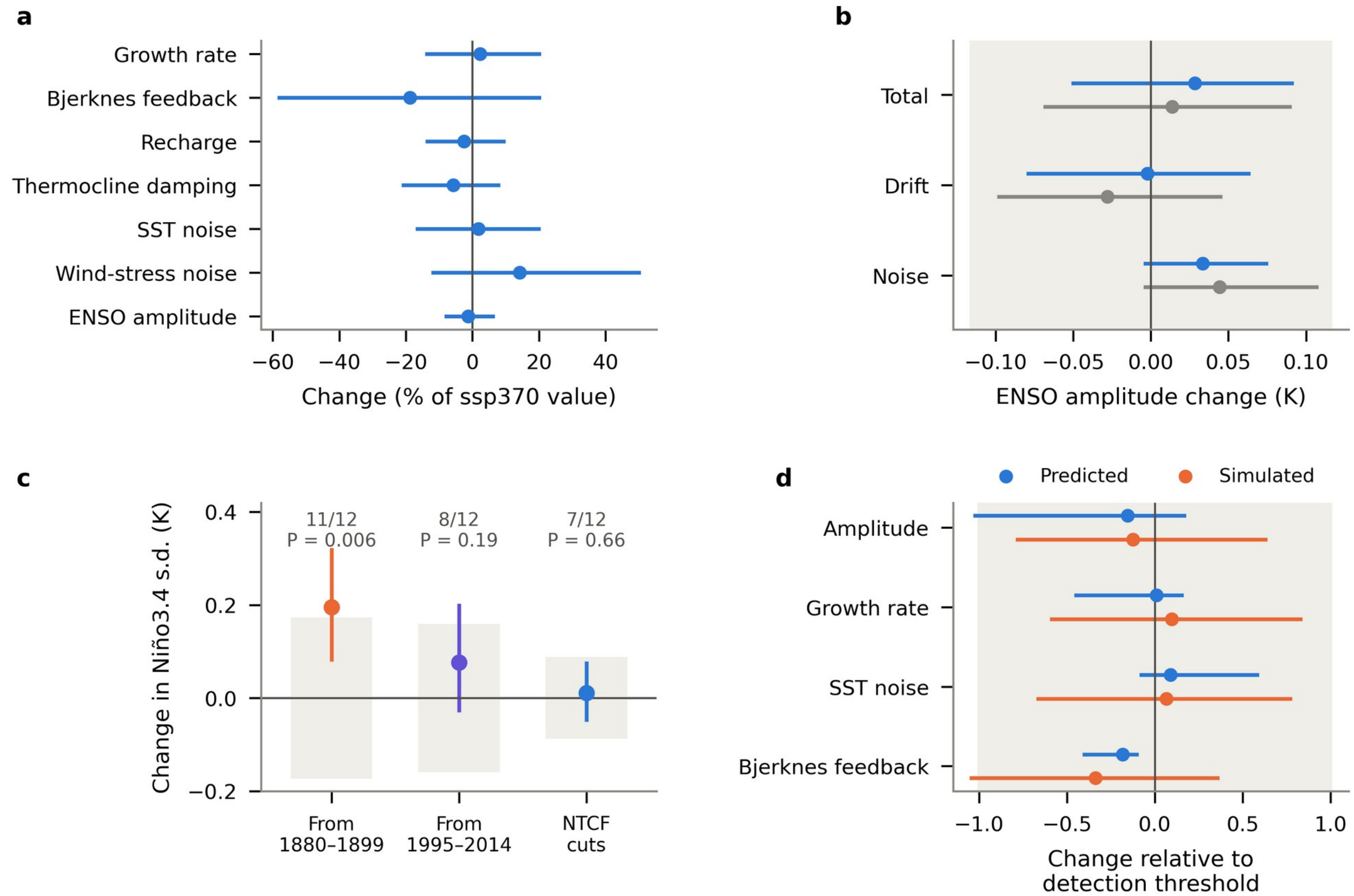
a
Growth rate
Bjerknes feedback
Recharge
Thermocline damping
SST noise
Wind-stress noise
ENSO amplitude
−60 −40 −20 0 20 40
Change (% of ssp370 value)
b
Total
Drift
Noise
−0.10 −0.05 0.00 0.05 0.10
ENSO amplitude change (K)
c
Change in Niño3.4 s.d. (K)
11/12
P = 0.006
8/12
P = 0.19
7/12
P = 0.66
From 1880–1899
From 1995–2014
NTCF cuts
d
Predicted
Simulated
Amplitude
Growth rate
SST noise
Bjerknes feedback
Change relative to detection threshold

Fig. 5

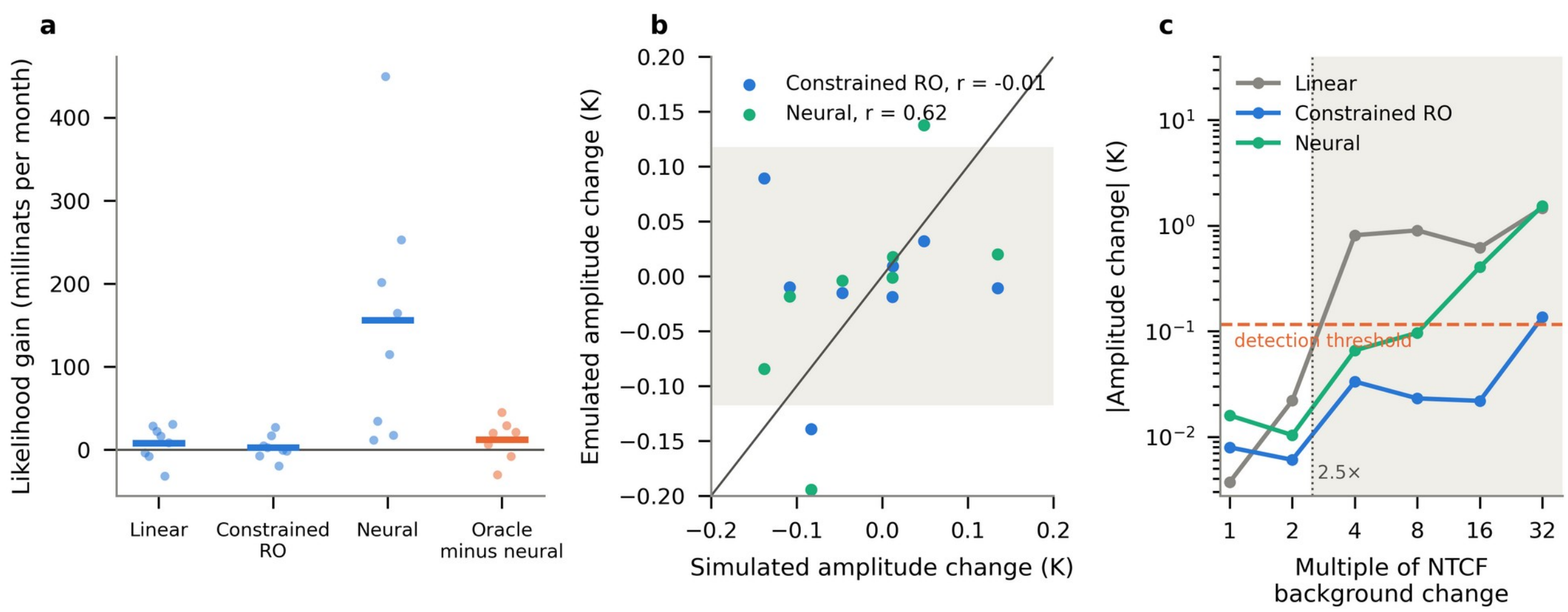
a
Likelihood gain (millinats per month)
Linear
Constrained RO
Neural
Oracle minus neural
b
Constrained RO, r = -0.01
Neural, r = 0.62
Emulated amplitude change (K)
Simulated amplitude change (K)
c
Linear
Constrained RO
Neural
|Amplitude change| (K)
detection threshold
2.5×
Multiple of NTCF background change

Fig. 6

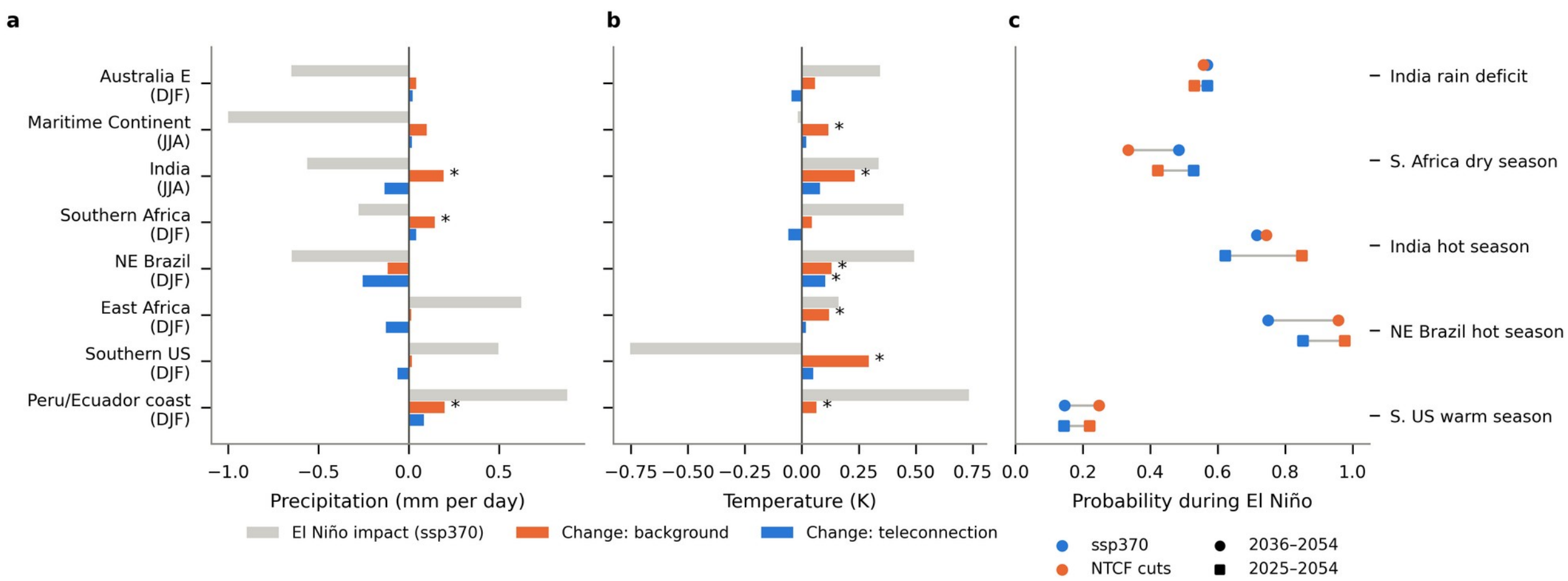

# Extended Data figures

## Extended Data Fig. 1

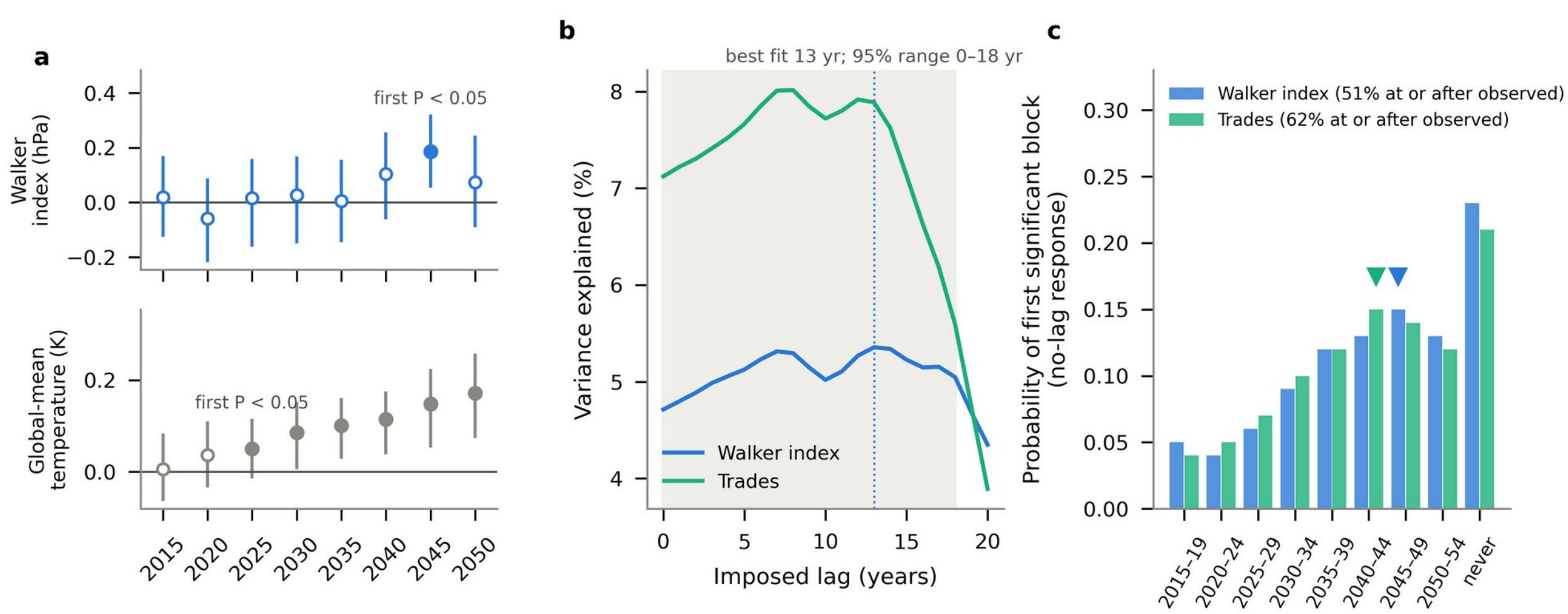

## Extended Data Fig. 2

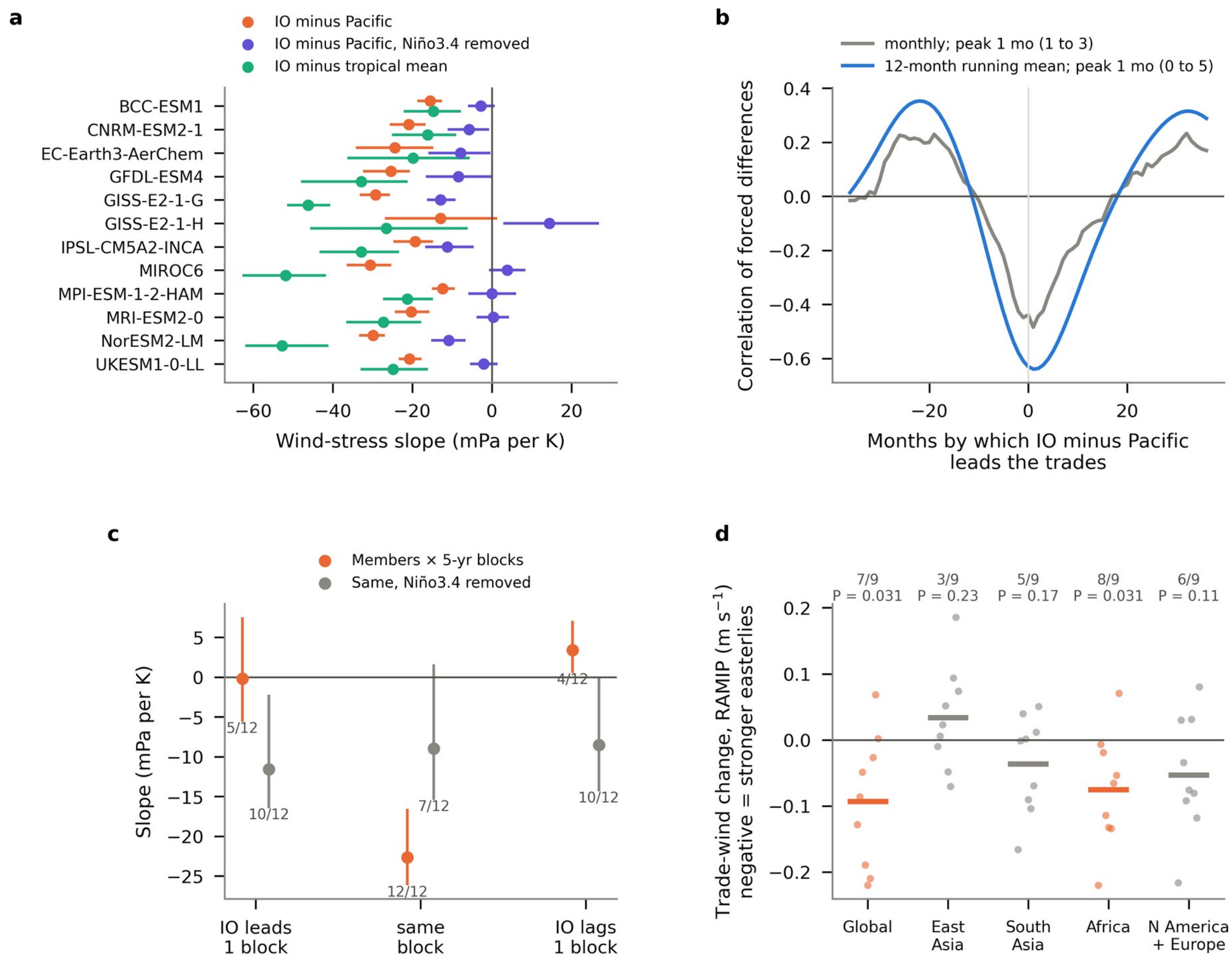

**Extended Data Fig. 3**

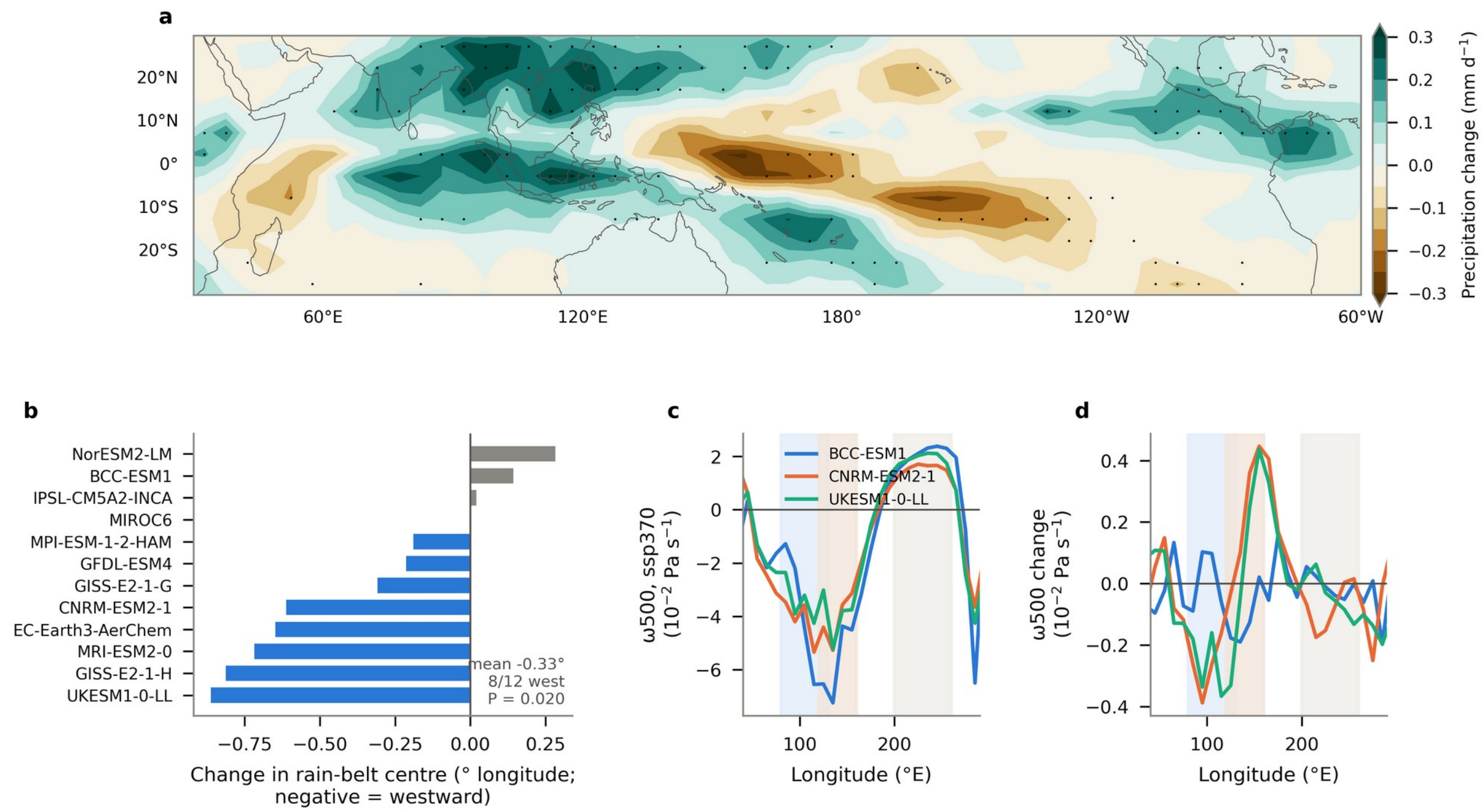

# Extended Data Fig. 4

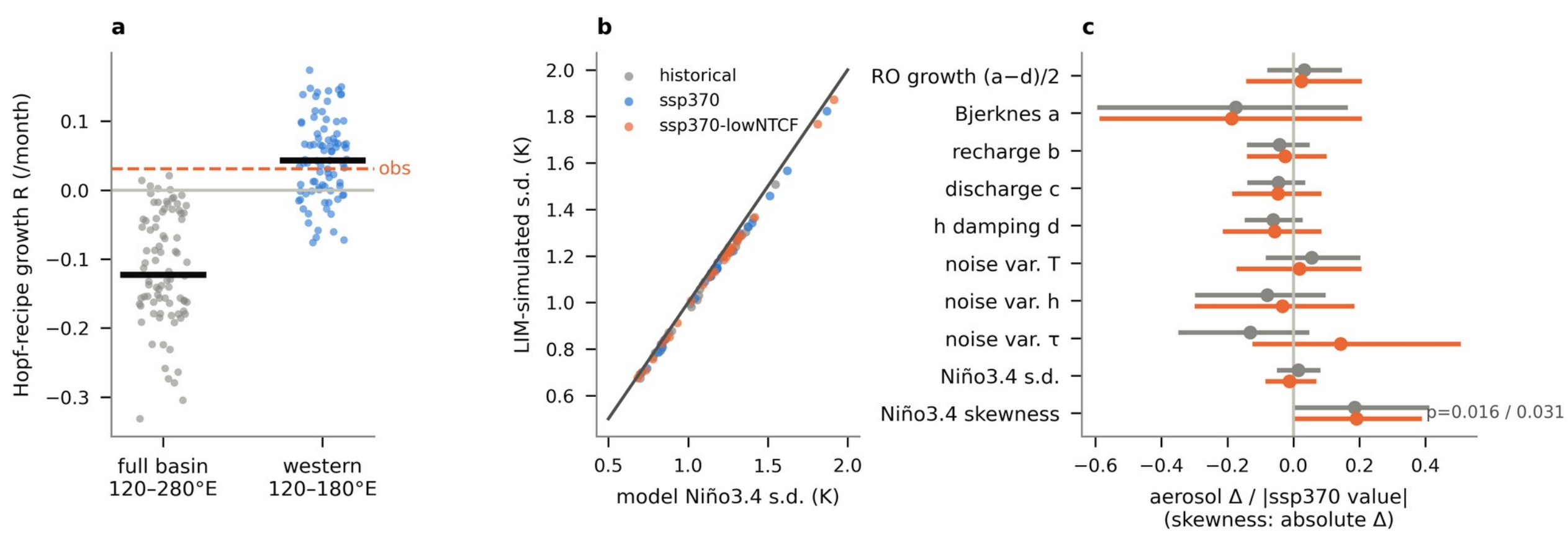

# Extended Data Fig. 5

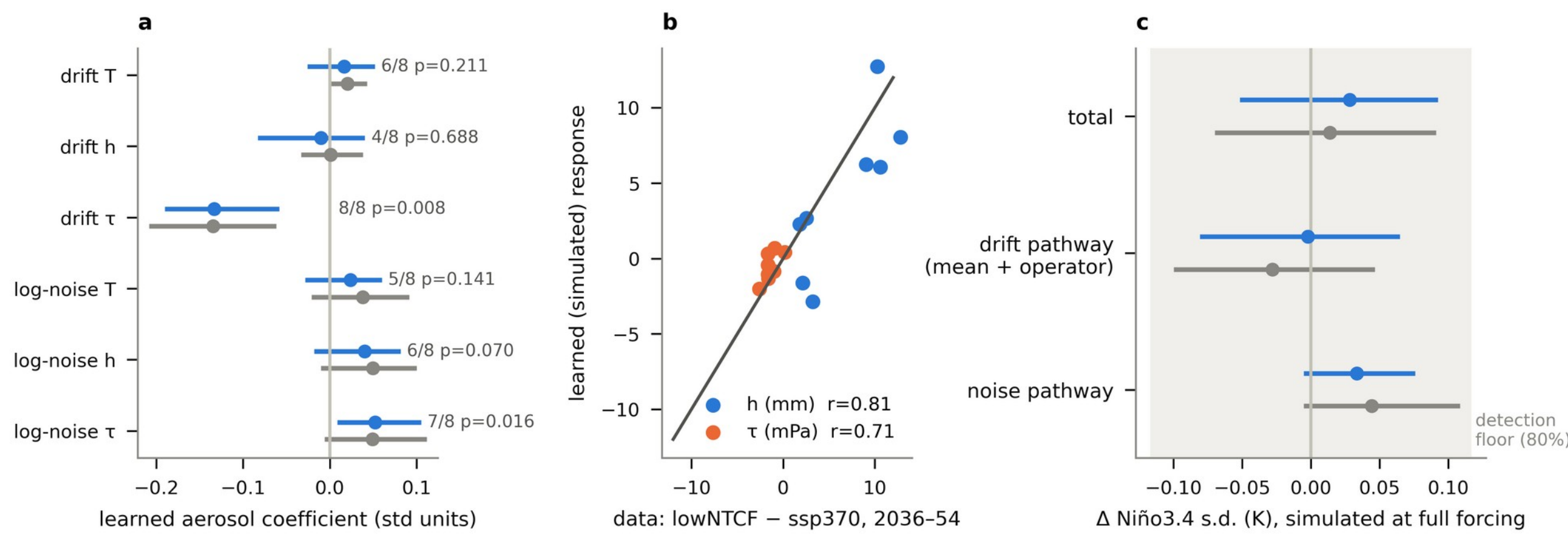

# Extended Data Fig. 6

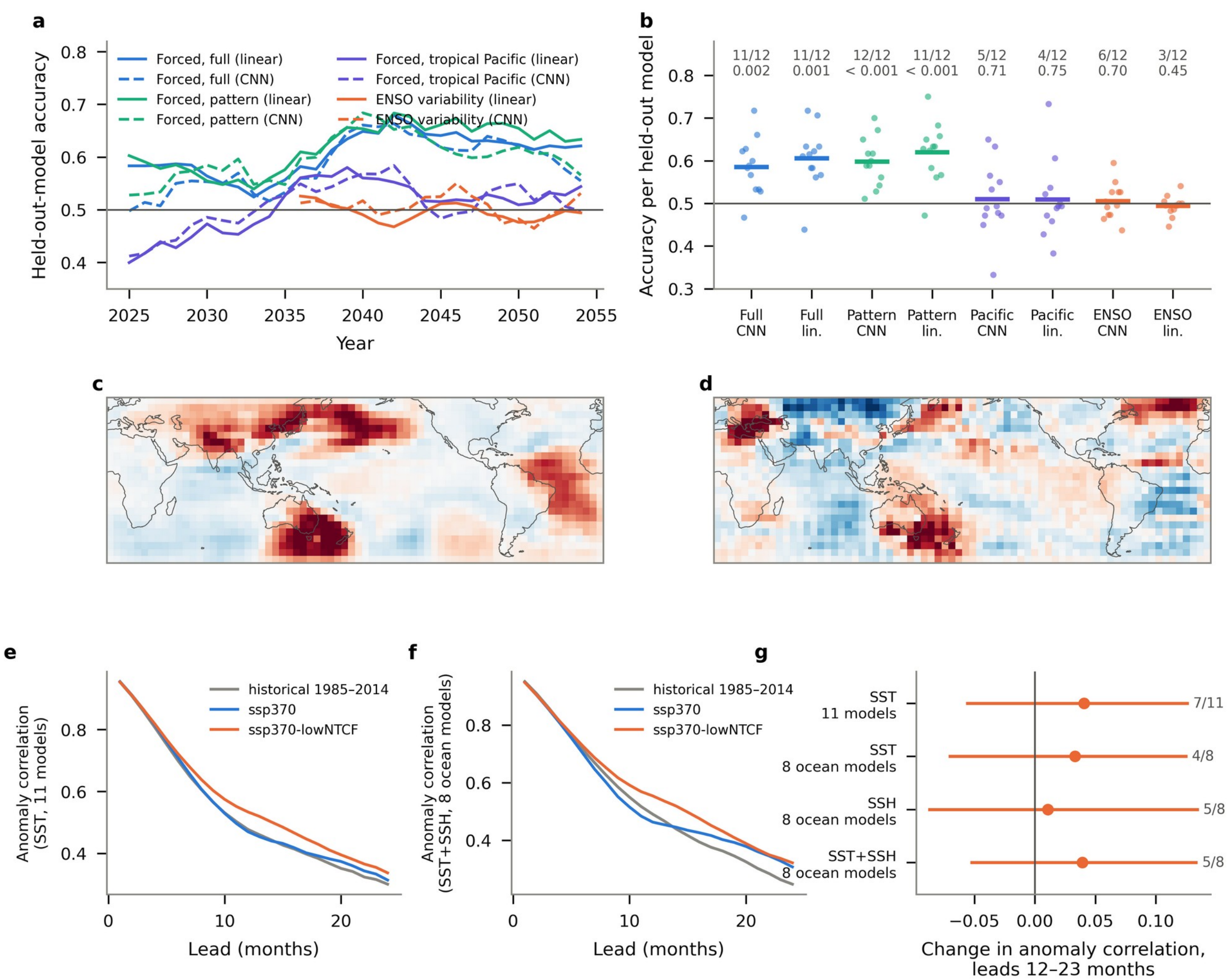

# Supplementary Information

## Cleaner air strengthens the Pacific trade winds without a detectable change in El Niño

Ioana Colfescu[1,2,3]*, Alice Pigelet[3], Chen Lu[1,2,3], Massimo A. Bollasina[4], Silvio Gualdi[5]

[1]National Centre for Atmospheric Science, University of St Andrews, St Andrews, UK

[2]School of Earth and Environmental Sciences, University of St Andrews, St Andrews, UK

[3]School of Mathematics and Statistics, University of St Andrews, St Andrews, UK

[4]School of GeoSciences, University of Edinburgh, Edinburgh, UK

[5]Fondazione Centro Euro-Mediterraneo sui Cambiamenti Climatici (CMCC), Bologna, Italy

*e-mail: ioana.colfescu@ncas.ac.uk

### Contents

All experiments, indices, windows and tests are defined in the Methods of the main text. P values are from the exact two-sided model-level sign-flip test unless the legend says otherwise.

**Supplementary Table 1 | Neural SDE placebo and held-out tests by model.**

| Model | Assignments | Trade forcing, real | Trade forcing, null mean | P (model) | Drift-only change in s.d. (K), real | P (model) | Held-out gain |
|---|---|---|---|---|---|---|---|
| BCC-ESM1 | 20 | −0.0848 | 0.0143 | 0.65 | −0.0173 | 0.75 | −0.00522 |
| CNRM-ESM2-1 | 20 | −0.136 | −0.00893 | 0.15 | 0.0938 | 0.2 | −0.0109 |
| EC-Earth3-AerChem | 20 | −0.241 | 0.0237 | 0.1 | 0.0143 | 0.9 | 0.023 |
| GISS-E2-1-G | 201 | −0.0476 | −0.00188 | 0.249 | −0.0313 | 0.801 | 0.0211 |
| GISS-E2-1-H | 6 | −0.0207 | 0.0348 | 0.833 | −0.0694 | 0.667 | −0.0526 |
| MPI-ESM-1-2-HAM | 20 | −0.208 | 0.0729 | 0.25 | 0.0195 | 0.95 | −0.00301 |
| MRI-ESM2-0 | 20 | −0.118 | 0.0112 | 0.25 | 0.0164 | 1 | 0.00766 |
| NorESM2-LM | 6 | −0.104 | −0.00306 | 0.167 | −0.107 | 0.833 | 0.0208 |
| 8-model mean | 313 | −0.12 | 0.0178 | 0.00318 | −0.0101 | 0.811 | 0.0001 |

For each model we reassigned the members between ssp370 and ssp370-lowNTCF, keeping the number in each experiment fixed, and refitted the neural SDE for every distinct assignment: 20 assignments for three members per experiment, 6 for two, and 200 random assignments for GISS-E2-1-G. The per-model P is the fraction of assignments with a trade-forcing coefficient at least as extreme as the real one, so the smallest values attainable are 0.05 and 0.17. The 8-model P compares the real mean with 200,000 means formed by drawing one placebo fit per model, P = (1 + count)/(1 + 200,000). The trade-forcing coefficient is in standardised units per month. The drift-only change is the change in the Niño3.4 standard deviation when the learned forcing acts in the drift alone, with its own placebo P. The held-out gain is the likelihood gain, on a withheld ssp370-lowNTCF member, of an SDE that sees the forcing over one that does not, averaged over the withheld members; a gain near zero means that knowing the forcing does not help the model predict a member it has not seen.

**Supplementary Table 2 | El Niño impact terms with a false-discovery-rate adjustment.**

| Region (season) | Variable | bg 2036–2054 | P | q | tc 2036–2054 | P | q | bg 2025–2054 | P | q | tc 2025–2054 | P | q |
|---|---|---|---|---|---|---|---|---|---|---|---|---|---|
| Australia E (DJF) | precipitation | +0.042 | 0.454 | 0.518 | +0.021 | 0.797 | 0.904 | +0.035 | 0.377 | 0.503 | +0.018 | 0.788 | 0.846 |
| Australia E (DJF) | temperature | +0.057 | 0.071 | 0.104 | −0.046 | 0.468 | 0.716 | +0.029 | 0.329 | 0.478 | −0.050 | 0.315 | 0.630 |
| Maritime Continent (JJA) | precipitation | +0.100 | 0.052 | 0.084 | +0.017 | 0.848 | 0.904 | +0.088 | 0.069 | 0.138 | −0.015 | 0.793 | 0.846 |
| Maritime Continent (JJA) | temperature | +0.116 | 0.001 | 0.008 | +0.020 | 0.193 | 0.603 | +0.086 | 0.004 | 0.016 | +0.003 | 0.714 | 0.846 |
| India (JJA) | precipitation | +0.193 | 0.021 | 0.042 | −0.135 | 0.428 | 0.716 | +0.117 | 0.029 | 0.077 | −0.040 | 0.665 | 0.846 |
| India (JJA) | temperature | +0.231 | 0.000 | 0.008 | +0.079 | 0.388 | 0.716 | +0.163 | 0.000 | 0.008 | +0.023 | 0.552 | 0.846 |
| Southern Africa (DJF) | precipitation | +0.144 | 0.017 | 0.042 | +0.042 | 0.496 | 0.716 | +0.085 | 0.079 | 0.140 | +0.073 | 0.062 | 0.247 |
| Southern Africa (DJF) | temperature | +0.044 | 0.435 | 0.518 | −0.060 | 0.226 | 0.603 | +0.037 | 0.438 | 0.538 | −0.049 | 0.142 | 0.379 |
| NE Brazil (DJF) | precipitation | −0.116 | 0.319 | 0.425 | −0.255 | 0.088 | 0.603 | −0.035 | 0.622 | 0.710 | −0.117 | 0.280 | 0.630 |
| NE Brazil (DJF) | temperature | +0.130 | 0.011 | 0.034 | +0.102 | 0.006 | 0.094 | +0.096 | 0.006 | 0.020 | +0.056 | 0.077 | 0.247 |
| East Africa (DJF) | precipitation | +0.014 | 0.848 | 0.848 | −0.126 | 0.172 | 0.603 | +0.002 | 0.956 | 0.956 | −0.124 | 0.057 | 0.247 |
| East Africa (DJF) | temperature | +0.120 | 0.004 | 0.016 | +0.017 | 0.582 | 0.716 | +0.089 | 0.003 | 0.016 | +0.006 | 0.891 | 0.891 |
| Southern US (DJF) | precipitation | +0.018 | 0.684 | 0.729 | −0.061 | 0.128 | 0.603 | +0.002 | 0.930 | 0.956 | −0.078 | 0.041 | 0.247 |
| Southern US (DJF) | temperature | +0.293 | 0.001 | 0.008 | +0.051 | 0.557 | 0.716 | +0.221 | 0.002 | 0.016 | −0.029 | 0.716 | 0.846 |
| Peru/ Ecuador coast (DJF) | precipitation | +0.199 | 0.019 | 0.042 | +0.084 | 0.397 | 0.716 | +0.104 | 0.163 | 0.260 | +0.157 | 0.014 | 0.227 |
| Peru/ Ecuador coast (DJF) | temperature | +0.064 | 0.034 | 0.061 | +0.001 | 0.989 | 0.989 | +0.061 | 0.039 | 0.088 | −0.020 | 0.374 | 0.665 |

Change under the NTCF cuts (ssp370-lowNTCF minus ssp370) in the background term (bg) and the teleconnection term (tc) of the impact of an El Niño of 1.5 K, mean of 12 models, for the 2036–2054 and 2025–2054 windows; precipitation is in mm per day and temperature in K. The background term is the change in the seasonal mean itself; the teleconnection term is the change in the regression of the seasonal anomaly on the Niño3.4 index, scaled to 1.5 K. DJF is December to February and JJA June to August. P is from the exact model-level sign-flip test. q is the Benjamini–Hochberg adjusted value over the 16 region–variable combinations within each window and term. None of the teleconnection terms has $q < 0.05$ in either window. In 2036–2054, 8 of the 16 background terms do. Adjusted over all 64 tests instead, the temperature background terms for India, the Maritime Continent, East Africa and the southern United States keep $q < 0.05$ in both windows, the India and southern Africa rainfall terms do not ($q = 0.09$ and 0.08), and the north-eastern Brazil temperature teleconnection term does ($q = 0.04$). The deposited file elnino_impacts_fdr.csv also gives the adjustment over all 64 tests.

**Supplementary Table 3 | Observed relation between the Indian Ocean minus Pacific temperature and the western-Pacific zonal wind.**

| SST dataset | Period | Niño3.4 removed | Trend removed | r (annual) | P (annual) | n (years) | r (5-year means) | P (5-year) | n (blocks) | Slope (m $s^{-1}$ per K) |
|---|---|---|---|---|---|---|---|---|---|---|
| ESA-CCI v2.1 | 1982–2017 | no | no | −0.77 | < 0.001 | 36 | −0.93 | 0.004 | 7 | −3.5 |
| ESA-CCI v2.1 | 1982–2017 | no | yes | −0.76 | < 0.001 | 36 | −0.89 | 0.010 | 7 | −3.6 |
| ESA-CCI v2.1 | 1982–2017 | yes | no | −0.53 | 0.006 | 36 | −0.81 | 0.027 | 7 | −2.6 |
| ESA-CCI v2.1 | 1982–2017 | yes | yes | −0.33 | 0.069 | 36 | −0.55 | 0.206 | 7 | −1.4 |
| HadISST 1.1 | 1980–2014 | no | no | −0.81 | < 0.001 | 35 | −0.95 | 0.002 | 7 | −4.9 |
| HadISST 1.1 | 1980–2014 | no | yes | −0.78 | < 0.001 | 35 | −0.93 | < 0.001 | 7 | −5.4 |
| HadISST 1.1 | 1980–2014 | yes | no | −0.55 | < 0.001 | 35 | −0.87 | 0.013 | 7 | −3.4 |
| HadISST 1.1 | 1980–2014 | yes | yes | −0.41 | 0.007 | 35 | −0.74 | 0.057 | 7 | −3.4 |

The temperature difference is the tropical Indian Ocean minus the tropical Pacific, defined as in the models, from HadISST 1.1 (1980–2014) or ESA-CCI SST v2.1 (1982–2017). The wind is the ERA5 zonal wind at 1000 hPa averaged over the western-Pacific box, because monthly surface stress is not held on the JASMIN archive; a negative r means that a warmer Indian Ocean relative to the Pacific goes with stronger easterlies. Niño3.4 is removed by regression from both series where indicated, and the trend by a linear fit. For the annual series P is from 10,000 shuffles of 5-year blocks; for the 5-year means it is exact over all orderings of the 7 blocks. The slope is for the 5-year means. The Discussion cites the rows with both Niño3.4 and the trend removed.

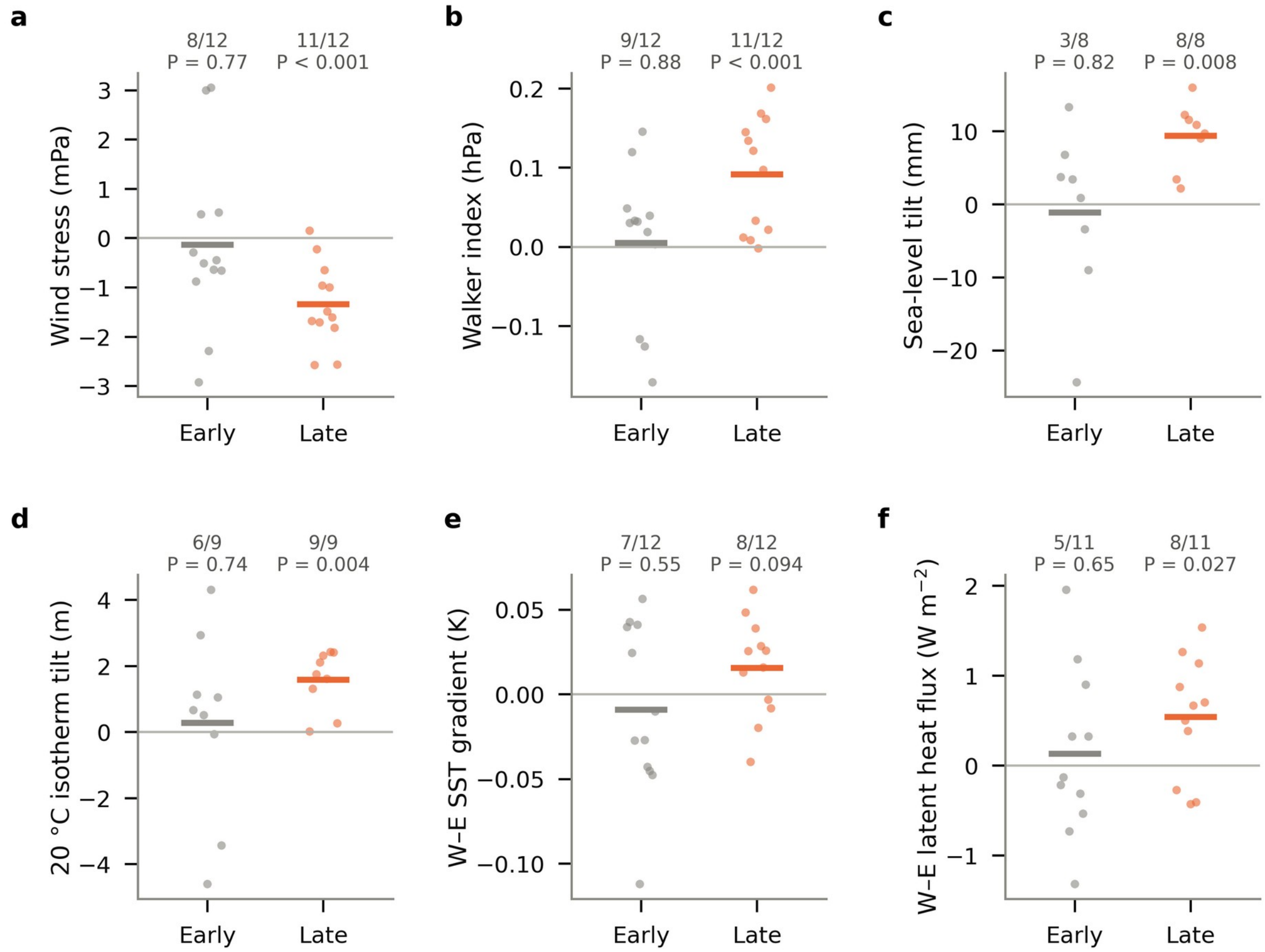

**Supplementary Fig. 1 | Growth test for six mean-state quantities.** Per-model differences between ssp370-lowNTCF and ssp370 in the early (2015–2035) and late (2036–2054; 2036–2055 for the ocean fields) windows for western-Pacific wind stress (a), the Walker index (b), sea-level tilt (c), 20 °C isotherm tilt (d), the west-minus-east SST gradient (e) and the west-minus-east latent heat flux (f). Dots are models and bars their mean. The numbers give the models agreeing with the sign of the mean and P from the exact two-sided sign-flip test (n = 12, 12, 8, 9, 12 and 11 models). Each quantity is larger and more consistent in the late window than in the early one, which is the growth test referred to in the main text.

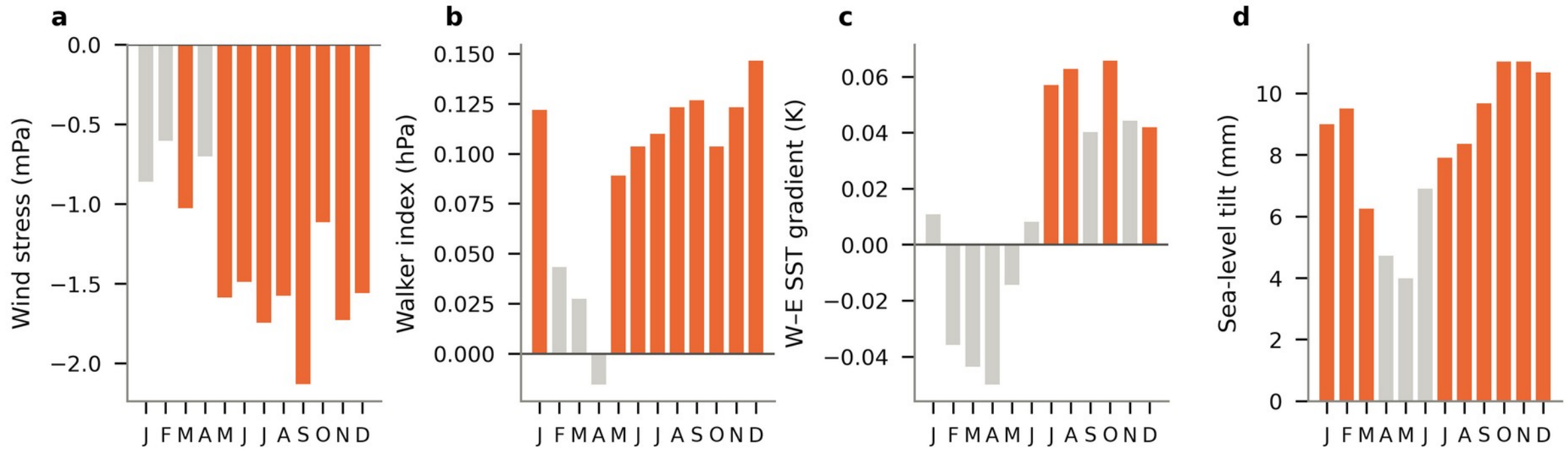

**Supplementary Fig. 2 | Seasonality of the mean-state response.** Calendar-month difference between ssp370-lowNTCF and ssp370 over 2036–2054 in western-Pacific wind stress (a), the Walker index (b), the west-minus-east SST gradient (c) and the sea-level tilt (d), mean of 12 models (8 for d). Orange bars are significant at $P < 0.05$ in the sign-flip test and grey bars are not. The trades and the pressure gradient strengthen from May to December and change little from February to April.

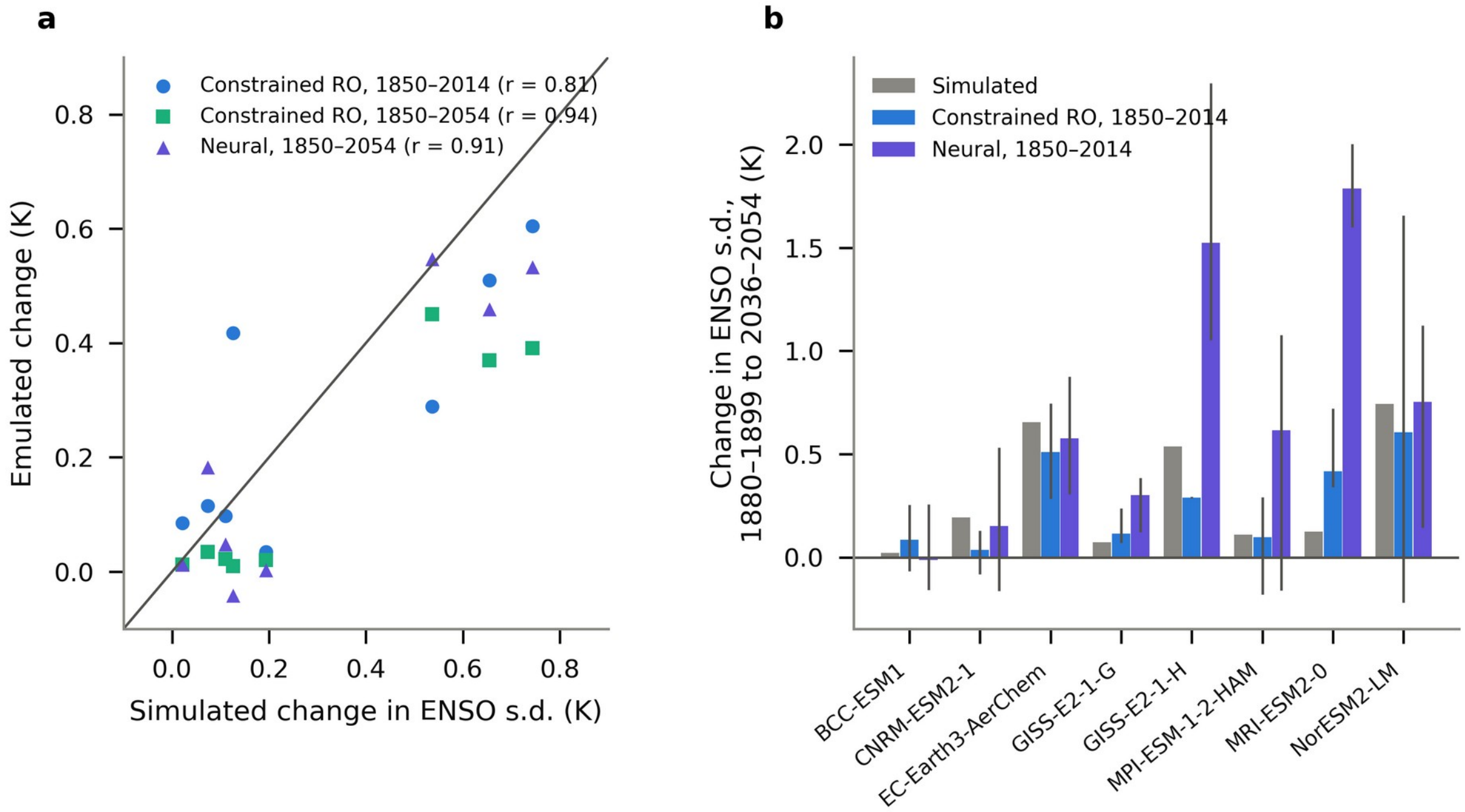

**Supplementary Fig. 3 | Emulator positive control per model.** a, Emulated against simulated change in ENSO amplitude between the 1880–1899 background of the historical runs and the 2036–2054 background of ssp370, for the constrained recharge oscillator trained on 1850–2014 and on 1850–2054 and for the neural emulator trained on 1850–2054; r is the across-model correlation and the line is 1:1. b, Simulated and emulated change per model for the two levels trained on 1850–2014 only; whiskers span five member-bootstrap refits. Amplitude is the standard deviation of Niño3.4 anomalies about the 21-year running climatology (n = 8 models). The constrained level reproduces the greenhouse-driven rise from the historical period alone. The neural level does not.